\documentclass[preprint,journal]{vgtc}            % preprint (journal style)
\onlineid{9953}

\ieeedoi{10.1109/TVCG.2026.3694447}

\vgtccategory{Research}

\vgtcpapertype{algorithm/technique}

\title{Enhancing Line Density Plots with Outlier Control\\and Bin-based Illumination}

\author{%
  \authororcid{Yumeng Xue}{0000-0002-8195-517X},
  \authororcid{Bin Chen}{0009-0005-5047-214X},
  \authororcid{Patrick Paetzold}{0000-0002-1315-4602}, \authororcid{Yunhai Wang}{0000-0003-0059-6580}, \authororcid{Christophe Hurter}{0000-0003-4318-6717}, and \authororcid{Oliver Deussen}{0000-0001-5803-2185}
}

\authorfooter{
  \item
  	Y. Xue, B. Chen, P. Paetzold, and O. Deussen are with University of Konstanz, Germany.
  	E-mail: \{ﬁrstname.lastname\}@uni-konstanz.de.
  \item
  	Y. Wang is with Renmin University of China, China.
  	E-mail: wang.yh@ruc.edu.cn.
  \item C. Hurter is with Fédération ENAC ISAE-SUPAERO ONERA, Université de Toulouse, France, and with IPAL Singapore International Research Laboratory on Artificial Intelligence, Singapore 138632
  	E-mail: christophe.hurter@enac.fr.
}

\abstract{
    Density plots effectively summarize large numbers of points, which would otherwise lead to severe overplotting in, for example, a scatter plot. However, when applied to line-based datasets, such as trajectories or time series, density plots alone are insufficient, as they disrupt path continuity, obscuring smooth trends and rare anomalies. We propose a bin-based illumination model that decouples structure from density to enhance flow and reveal sparse outliers while preserving the original colormap. We introduce a bin-based outlierness metric to rank trajectories. Guided by this ranking, we construct a structural normal map and apply locally-adaptive lighting in the luminance channel to highlight chosen patterns---from dominant trends to atypical paths---with acceptable color distortion. Our interactive method enables analysts to prioritize main trends, focus on outliers, or strike a balance between the two. We demonstrate our method on several real-world datasets, showing it reveals details missed by simpler alternatives, achieves significantly lower CIEDE2000 color distortion than standard shading, and supports interactive updates for up to 10,000 lines.
}


 \teaser{
   \centering
    \begin{subfigure}[b]{0.32\textwidth}
        \centering
        \includegraphics[width=\textwidth]{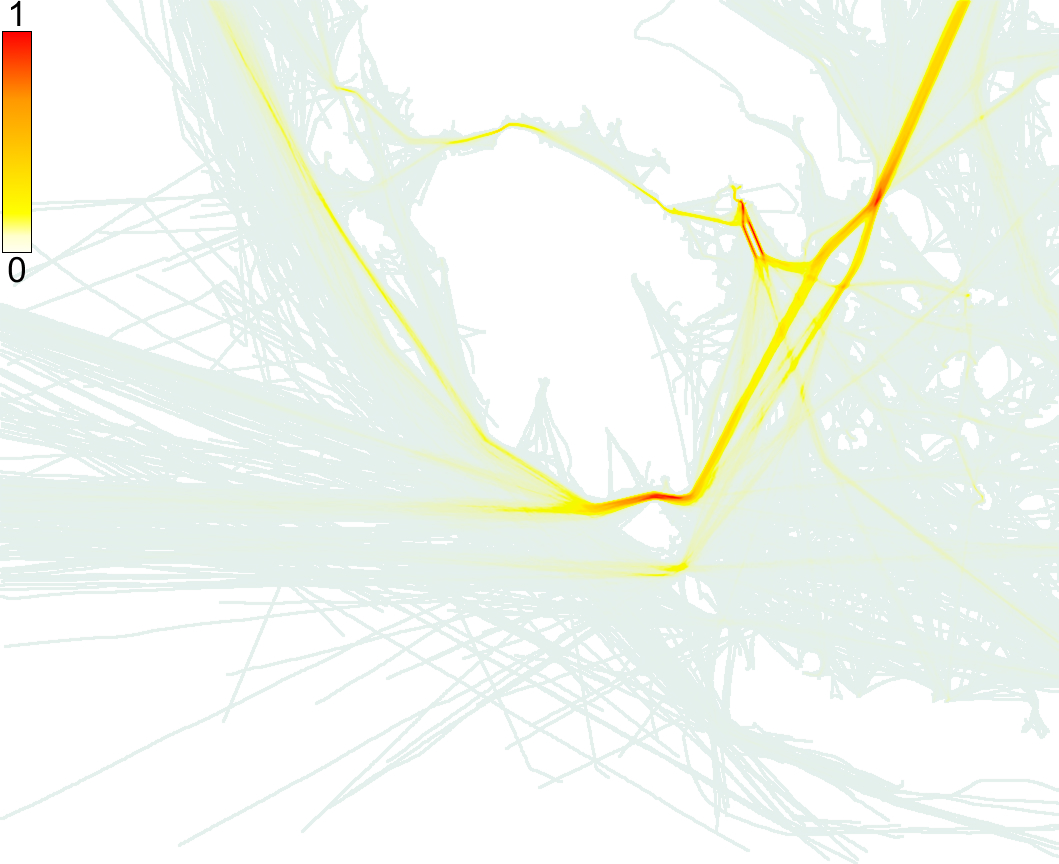}
        \caption{Plain density plot}
        \label{fig:teaser:a}
    \end{subfigure}
    \hfill
    \begin{subfigure}[b]{0.32\textwidth}
        \centering
        \includegraphics[width=\textwidth]{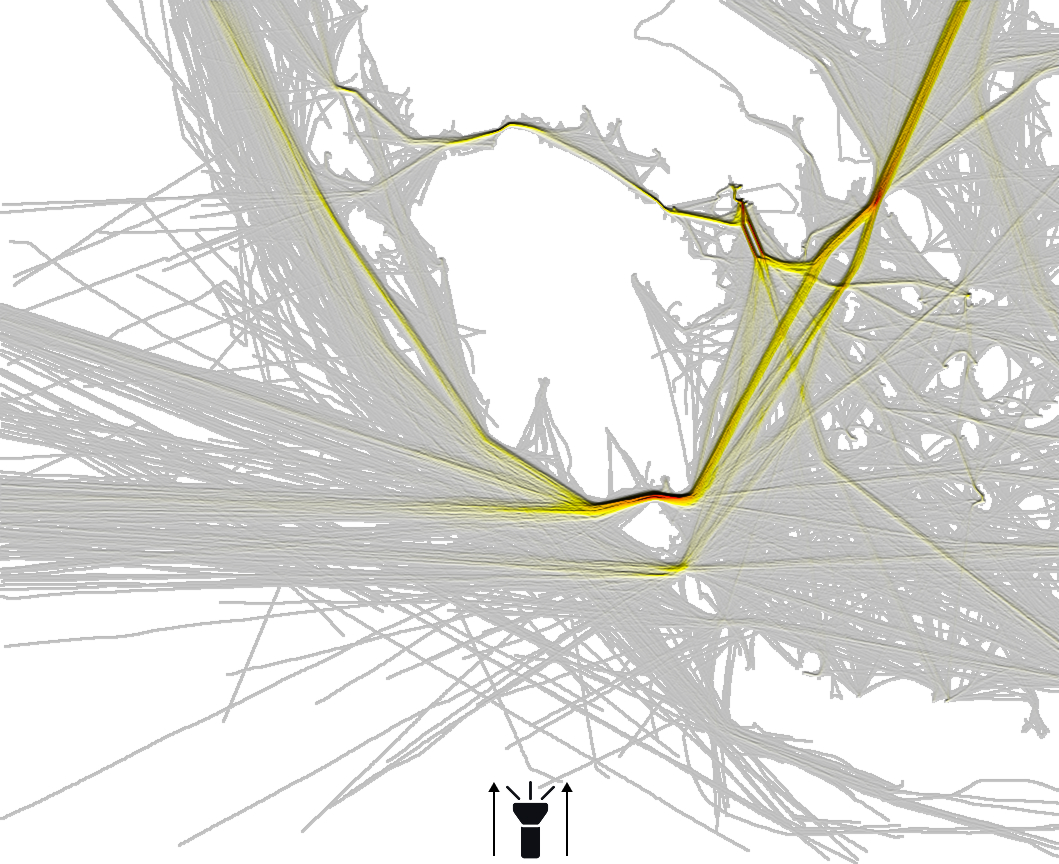}
        \caption{Direct Lambertian shading in RGB}
        \label{fig:teaser:b}
    \end{subfigure}
    \hfill
    \begin{subfigure}[b]{0.32\textwidth}
        \centering
        \includegraphics[width=\textwidth]{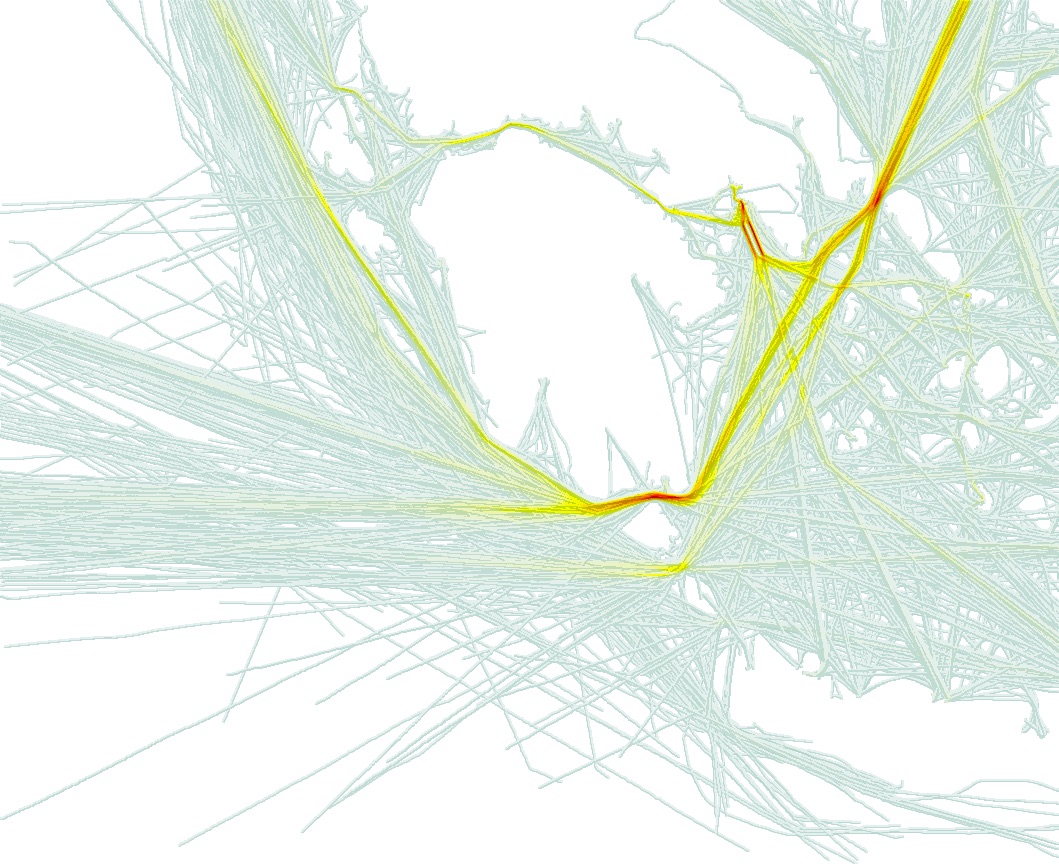}
        \caption{Our bin-based illumination method}
        \label{fig:teaser:c}
    \end{subfigure}
    \vspace{-3mm}
   \caption{%
    Comparison of line-density visualization methods on vessel trajectories. 
    (a) Plain density plot highlights only high-density regions while obscuring continuity and sparse lines;
    (b) Direct Lambertian shading in RGB introduces color distortion and orientation bias;
    (c) Our bin-based illumination decouples normals from density, adapts light per bin, and preserves the colormap while enhancing both dense trends and sparse anomalies.}
   \label{fig:teaser}
}

\renewcommand{\manuscriptnotetxt}{}

\graphicspath{{figs/}{figures/}{pictures/}{images/}{./}} % where to search for the images

\usepackage{tabu}                      % only used for the table example
\usepackage{booktabs}                  % only used for the table example
\usepackage{lipsum}                    % used to generate placeholder text
\usepackage{mwe}                       % used to generate placeholder figures

\usepackage{wrapfig}

\usepackage{amsfonts}

\usepackage{mathptmx}                  % use matching math font
\usepackage{amsmath}
\usepackage{multirow}
\usepackage{pifont}

\begin{document}

%%%%%%%%%%%%%%%%%%%%%%%%%%%%%%%%%%%%%%%%%%%%%%%%%%%%%%%%%%%%%%%%
%%%%%%%%%%%%%%%%%%%%%% START OF THE PAPER %%%%%%%%%%%%%%%%%%%%%%
%%%%%%%%%%%%%%%%%%%%%%%%%%%%%%%%%%%%%%%%%%%%%%%%%%%%%%%%%%%%%%%%


\firstsection{Introduction}

\maketitle

Density plots are widely used to visualize large and complex datasets, particularly when traditional methods such as scatterplots suffer from overplotting and visual clutter. By aggregating data points into discrete bins and encoding their density through color gradients, density plots effectively reveal the global distribution and coarse structures of point-based data. However, this binning-based approach has notable limitations for line-based datasets (e.g., trajectories or time series), where continuity and local shape are essential: fixed bins can disrupt perceived flow, hide smooth trends, and obscure subtle structural differences between nearby trajectories. 
As illustrated in \cref{fig:teaser:a}, when visualizing vessel trajectories, only a few high-density patterns remain clearly visible, while sparse trajectories become barely visible, and the continuity of dense regions can only be inferred through Gestalt principles of continuity~\cite{ellis2013source} rather than through explicit line details.

Previous work has explored alternative enhancements to line density plots. For example, Xue~\textit{et al.}~\cite{xue2024reducing} proposed a color-based differentiation method to reduce ambiguity between different trends. Complementary to this, several approaches~\cite{scheepens2011composite, scheepens2011interactive, willems2009vessel} have attempted to address continuity loss and low-density invisibility by applying shading models such as Phong shading~\cite{phongshading}, typically reduced to Lambertian shading by omitting specular and ambient terms. However, these shading-based techniques suffer from three core limitations that impede their applicability to general line data.
First, previous line shading methods derive normal maps from gradients of the density field. Because these normals are tightly coupled to the density encoding, they rarely reveal structural information beyond what the density map already conveys, as they highlight main trends. Such trends are visually dominant in \cref{fig:teaser:b}.
Thus, fine structural details remain underrepresented.
Second, fixed global light directions create an orientation bias: lines nearly parallel to the light vector are much less emphasized than those orthogonal to the light direction (detailed in \cref{subsec:motivation}).
% In \cref{fig:teaser:b}, the lines coming from the top left are considerably less emphasized than those coming from the top right. 
In \cref{fig:teaser:b}, horizontal line segments are emphasized, while vertical ones appear flat and indistinct.
In addition, applying shading directly in RGB space produces chromatic shifts (color distortions) which interfere with the density values of the colormap, as seen in \cref{fig:teaser:b}.

Motivated by these observations, we propose the bin-based illumination approach shown in \cref{fig:teaser:c}. It advances the state of the art in three directions. First, we define a bin-based \emph{outlierness} measure that quantifies how much a trajectory’s tangent orientation deviates from those of its local neighbors; informally, parallel neighbors indicate high similarity and low outlierness, while crossing or shape-distinct neighbors indicate low similarity and high outlierness. This measure provides a basis for ranking trajectories and distinguishing between dominant trends and rare outliers. Second, guided by these scores, we construct a \emph{structural normal map} that combines trajectory-level cues with density-based gradients, decoupling normals from density and revealing structural details that density alone cannot capture. Because trajectories are ranked by outlierness, users can interactively adjust which patterns are emphasized by shifting the focus toward trend-conforming trajectories, strongly deviating ones, or any balance in between. 
Third, we adapt the illumination itself: instead of a fixed light source, we determine a local light direction per bin based on the dominant trajectory orientation and,
following Chen~\textit{et al.}~\cite{chen2024visualization}, apply shading primarily to the luminance channel of CIELAB to minimize color distortion.
Together, these components yield shading that enhances continuity in dense regions, makes sparse deviations more perceptible, and preserves the interpretability of the density colormap. As illustrated in \cref{fig:teaser:c}, our approach provides both enhanced structural visibility and flexibility of emphasis.

The key contributions of this work are:
\begin{itemize}
    \item A bin-based similarity metric designed for line data that ranks trajectories by their degree of outlierness, providing a foundation for smoothly adjustable emphasis on either high-density trends or low-density outliers
    \item A combined structural-normal and dynamic-lighting model, together with a luminance-only shading strategy, that enhances continuity and detail while preserving colors used in the density map
    \item An image-synthesis pipeline that enables users to balance trend visibility and anomaly perception
\end{itemize}
By bridging the gap between the discrete representation of density plots and the continuous character of trajectory data, our method provides a density-based overview while making local shape deviations and sparse anomalies more perceptible. 
We validate our approach through a comprehensive evaluation comparing it against simpler alternatives, quantitative analyses of its color fidelity and runtime, and diverse case studies on real-world datasets.
\section{Related Work}
\label{sec:related}

In this section, we survey density-based visualizations and discuss their application to continuous line data. We describe density estimation methods, ranging from foundational approaches to advanced techniques, while addressing challenges such as visual clutter and interpretability issues of complex datasets.
Additionally, we present our proposed illumination method.

\subsection{Density-based Visualization}
\label{subsec:related-density}
Density-based visualization techniques play a critical role in simplifying large datasets by reducing clutter, 
uncovering trends, and revealing distributions. Line-density plots have applications in time series and trajectory visualization across various domains, including finance~\cite{moritz2018visualizing, zhao2021kd} and transportation~\cite{willems2009vessel, hurter5290707}.

In early examples, sequential data of time series was aggregated by density visualizations to highlight value distributions over time. 
Work by Carr {\textit{et al.}}~\cite{carr1987scatterplot} introduced density overlays in scatterplot matrices to summarize large datasets. This approach is tailored to static point data rather than continuous temporal sequences. Kernel Density Estimation (KDE), introduced by Silverman~\cite{silverman1986density} and refined by Feng {\textit{et al.}}~\cite{feng2010matching}, is a foundational model to generate smooth density fields. Wickham~\cite{wickham2013bin} advanced this paradigm by employing binning, summarizing, and smoothing techniques, which map density to color scales to emphasize trends in large datasets. To simplify dense temporal data in line charts, Jerding and Stasko~\cite{jerding1998information} proposed to encode overlap levels with grayscale values.
However, these methods often prioritize density magnitude over temporal continuity, thereby obscuring the sequential relationships that are crucial for time series analysis.

For trajectory data, density-based methods aim to summarize spatial movement patterns. Scheepens {\textit{et al.}}~\cite{scheepens2011interactive} introduced adjustable KDE kernel radii for interactive exploration of trajectory density, later extending this with composite density maps that integrate attributes such as speed and direction~\cite{scheepens2011composite}.
Lampe {\textit{et al.}} proposed Curve Density Estimates (CDE)~\cite{lampe2011curve}, making KDE applicable to smooth curves and providing a robust framework for trajectory density visualization,
which we also use to generate our line density plot.
CDE defines a line segment using its endpoints \( p_i \) and \( p_{i+1} \), estimating density at a point \( x \) (using a Gaussian kernel \( N_{h} \)   with bandwidth \( h \), and the perpendicular projection \( w \) of \( x \) onto the segment) as:
\begin{wrapfigure}[6]{l}{20mm}
  \centering
  \includegraphics[width=22mm]{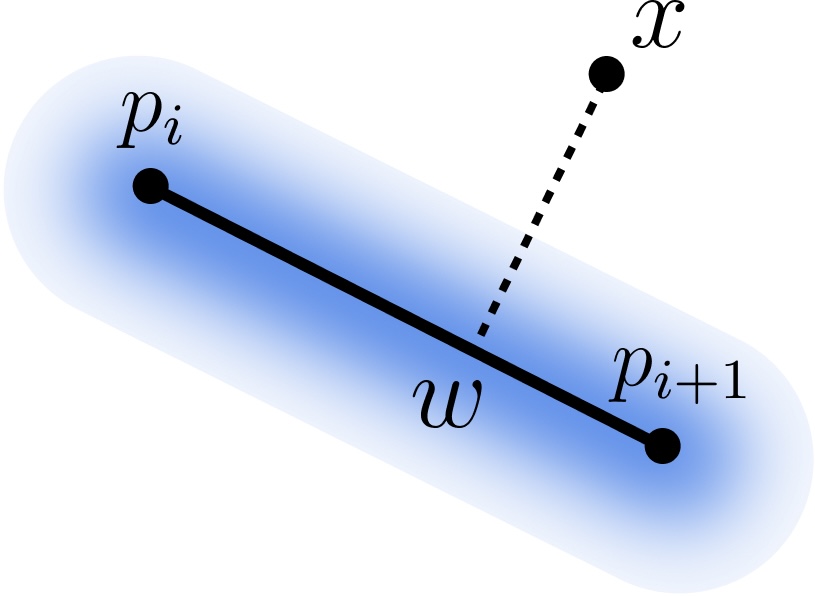}
  %\caption{CDE constructs a continuous density estimation between points of a line.}
\end{wrapfigure}
\begin{equation}
L_{h}(x, p_i, p_{i+1}) = L_{h}^{1D}(w, p_i, p_{i+1}) \cdot N_{h}(|x - w|)
\label{eq:line_kernel}
\end{equation}
The 1D line kernel is defined as:
\vspace{-1mm}
\begin{equation}
\hspace{10.5mm}
    L_h^{1D}(w, p_i, p_{i+1}) = \frac{\int_{p_i}^{p_{i+1}} N_h(w - t) \, \mathrm{d}t}{|p_{i+1} - p_i|}
    \label{eq:1D_line_kernel}
\vspace{-1mm}
\end{equation}
Applied to all segments of a curve, CDE yields a continuous density estimate that effectively captures line concentration.

Moritz and Fisher~\cite{moritz2018visualizing} discretized the computation of the continuous density estimate (CDE) for GPU execution with \textit{DenseLines}, enabling efficient visualization of large line datasets. However, their work did not provide a formal mathematical definition of a discretized CDE suitable for implementation. In this paper, we explicitly define a pixel-based discretization of CDE, which we use throughout our method. Specifically, we approximate the integral by discretizing the line segment. This is achieved by first rasterizing the segment into a set of pixels, \(P_{\text{segment}}\). The summation is then performed over these discrete pixel locations:
\vspace{-2mm}
\[
L_h^{1D}(w, p_i, p_{i+1}) \approx \frac{1}{|P_{\text{segment}}|} \sum_{p \in P_{\text{segment}}} N_h(w - p),
\vspace{-2mm}
\]
where \(p\) represents each pixel in the set \(P_{\text{segment}}\). We then place an isotropic Gaussian kernel at the center of each of these rasterized pixels, with bandwidth $h$ controlling the balance between smoothness and detail preservation.
This discretized formulation provides a concrete and reproducible bridge between the theoretical CDE and its practical implementation in our pipeline. 

\begin{figure*}[htb]
    \centering
    \includegraphics[width=\linewidth]{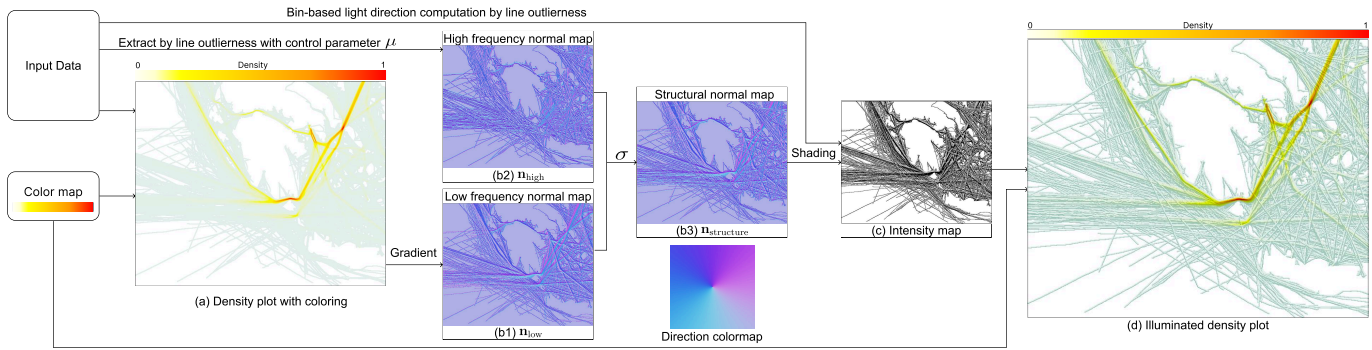}
    \vspace{-2mm}
    \caption{Pipeline for enhanced, discretized line density plots: (a) Initial density plot; (b) Normal maps, including low- and high-frequency normal maps, composed by user defined parameters to construct a structure normal map for rendering; (c) Intensity map with user-controlled low- and high-frequency patterns computed from structure map via bin-based light direction optimization; (d) Illuminated density plot combining density and shading.}
    \label{fig:pipeline}
    \vspace{-5mm}
\end{figure*}

\subsection{Density Plot Enhancement and Illumination}

Such enhancements encompass attribute adjustments as well as structural and interactive modifications. We will focus on introducing various illumination techniques that underpin our approach.

\noindent\textbf{Visual attribute adjustments} augment density maps by leveraging additional visual channels beyond coloring using colormaps. Matejka {\textit{et al.}}~\cite{matejka2015dynamic} proposed dynamic opacity blending to balance visibility between high- and low-density regions. To maintain visibility of outliers and distinguish cluster boundaries in dense areas, Splatterplots~\cite{Mayorga2013Splatterplots} display sub-sampled points explicitly in areas of low density and employ smooth, closed contours in regions with high density. Pomerenke {\textit{et al.}}~\cite{pomerenke2019slope} investigated how slope influences the perceived salience of lines within ghost clusters and introduced a density adjustment technique based on slope to minimize visual inaccuracies. For dense parallel coordinates, Novotný and Hauser~\cite{novotny2006outlier} developed a method to ensure that outliers remain visible. Micallef {\textit{et al.}}~\cite{micallef2017towards} advocate for a simultaneous optimization of all visual attributes, such as opacity, color, and mark sizes, to meet a set of manually defined perceptual indicators.
These approaches either modify the original density distribution or, by overlaying additional elements on the density plot, are in danger of obscuring the clarity of the density plot itself.

\noindent\textbf{Structural enhancements} analyze inherent data properties to refine density representations. Bao {\textit{et al.}}~\cite{bao2025bi} introduced a bi-scale framework with a variance-aware filter that adapts the smoothing process to preserve local structures in low-density regions while reducing over-smoothing in high-density regions. Topological methods~\cite{van2016multi} hold potential for extracting connectivity patterns within density fields, offering a mathematical basis for trend delineation. For geographic data visualizations, edge bundling techniques have been adapted to include constraints such as roads~\cite{thony2015vector, zeng2019route}, improving spatial coherence. Clustering-based approaches, such as pixel-based hierarchical clustering proposed by Xue~\textit{et al.}~\cite{xue2024reducing}, group regions by line similarity to reduce ambiguity, though they prioritize similarity over geometric continuity. Our proposed method can be combined with this approach to provide more details, see \cref{sec:evaluation:single-hue}.

\noindent\textbf{Interactive enhancements} like dynamic querying~\cite{hochheiser2004dynamic,mannino2018expressive}, brushing~\cite{zhao2021kd}, and multi-scale views~\cite{hurter5290707} enable users to navigate density maps interactively, mitigating the trade-off between overview and detail. We also employ interaction techniques for this purpose, see \cref{subsec:interaction}.

\noindent\textbf{Illumination-based techniques} emphasize geometric structures within density fields, enhancing visualizations beyond static colormaps through interaction. Willems {\textit{et al.}}~\cite{willems2009vessel} applied Phong shading to a height map derived from dual-bandwidth KDE, illuminating local structures. It alters the original colormap and struggles with low-density outliers. Trautner et al.~\cite{trautner2020sunspot} proposed Sunspot Plots, which combine shading with adaptive color blending to enhance the visibility of medium-density structures and outliers, though this comes at the expense of density value accuracy due to color distortions. Honeycomb Plots \cite{trautner2022honeycomb} further improve the shading to render outliers more distinguishable, but still face challenges in conveying local density variations.
More recently, Chen {\textit{et al.}}~\cite{chen2024visualization} proposed Visualization-Driven Illuminated Density Plots (VIDP), utilizing Difference of Gaussians (DoG) and luminance-only shading to enhance structure while preserving colormap integrity. However, VIDP targets point-based data; its filtering mechanisms neglect the sequential continuity and orientation inherent in trajectories. Consequently, it captures only local density variations, failing to visualize global flow or distinguish continuous trends from isolated outliers.
% More recently, Chen {\textit{et al.}}~\cite{chen2024visualization} proposed Visualization-Driven Illuminated Density Plots (VIDP), using a Difference of Gaussians (DoG) to compute a structure-enhancing shading field, paired with diffuse shading and optimized lighting to maximize contrast. By employing luminance-only composition, VIDP preserves the integrity of the colormap while revealing high-, medium-, and low-density structures, aligning illumination with visualization-specific goals rather than photorealistic rendering. 
% However, VIDP is tailored to point-based data; its point-based density calculation and DoG filtering operate on density fields generated by discrete points, and do not take into account continuity or orientation as present in line or trajectory data.
% When applied to line-based density plots, VIDP’s bin-wise enhancement mechanism neglects the intrinsic structure of trajectories—i.e., the fact that lines encode sequential data with inherent directionality and continuity. As a consequence, VIDP can at most highlight local density variations, but cannot recover or visualize the global flow of lines, nor distinguish between continuous trends and isolated outliers.
Our work extends these illumination techniques to prioritize line continuity within density plots, addressing the need for intuitive, unambiguous trend analysis and overcoming the limitations of prior discrete representations.
\section{Line Illumination Model and System}
\label{sec:method}

We present a bin-based illumination model that augments conventional line-density renderings with structure-aware shading. Our method operates on pixelized density fields and applies shading not only based on the density itself. We propose a pipeline (overview in \cref{fig:pipeline}) to compute trajectory-aware normals and locally adaptive lighting, and apply illumination only to the luminance channel to minimize chromatic interference. 
This pipeline serves a dual role: it summarizes our core algorithmic contributions and provides the foundational structure for the interactive system (detailed in \cref{subsec:interaction}). The pipeline's design is inseparably linked to our goal of real-time user interaction, as its key stages (e.g., structure emphasis and lighting) map directly to the parameters users can adjust to explore the data.
The following subsection motivates these design choices and outlines the goals that guided our pipeline design, followed by detailed technical specifications.

\subsection{Motivation, Overview, and Goals}
\label{subsec:motivation}

\begin{figure}[tb]
    \centering
    \hfill
    \begin{subfigure}{0.45\linewidth}
        \centering
        \includegraphics[width=\linewidth]{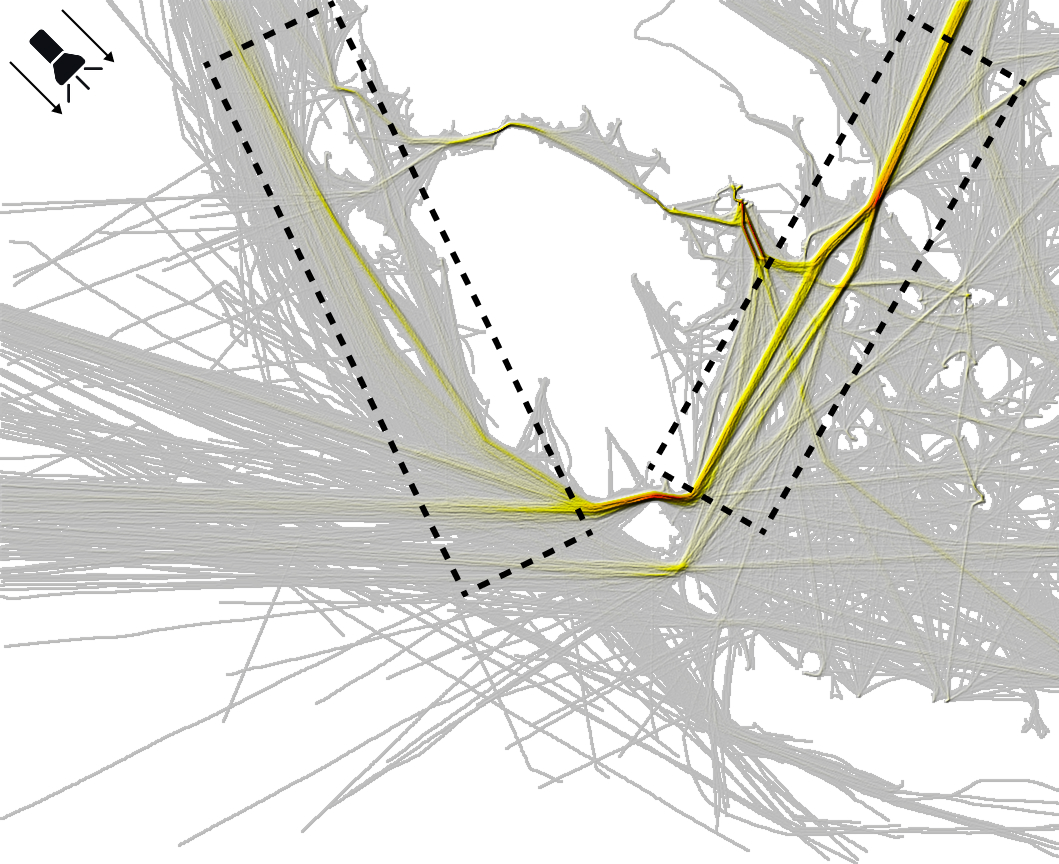}
        \caption{Lambertian shading with light from the northwest}
        \label{fig:left_top_light}
    \end{subfigure}
    \hfill
    \begin{subfigure}{0.45\linewidth}
        \centering
        \includegraphics[width=\linewidth]{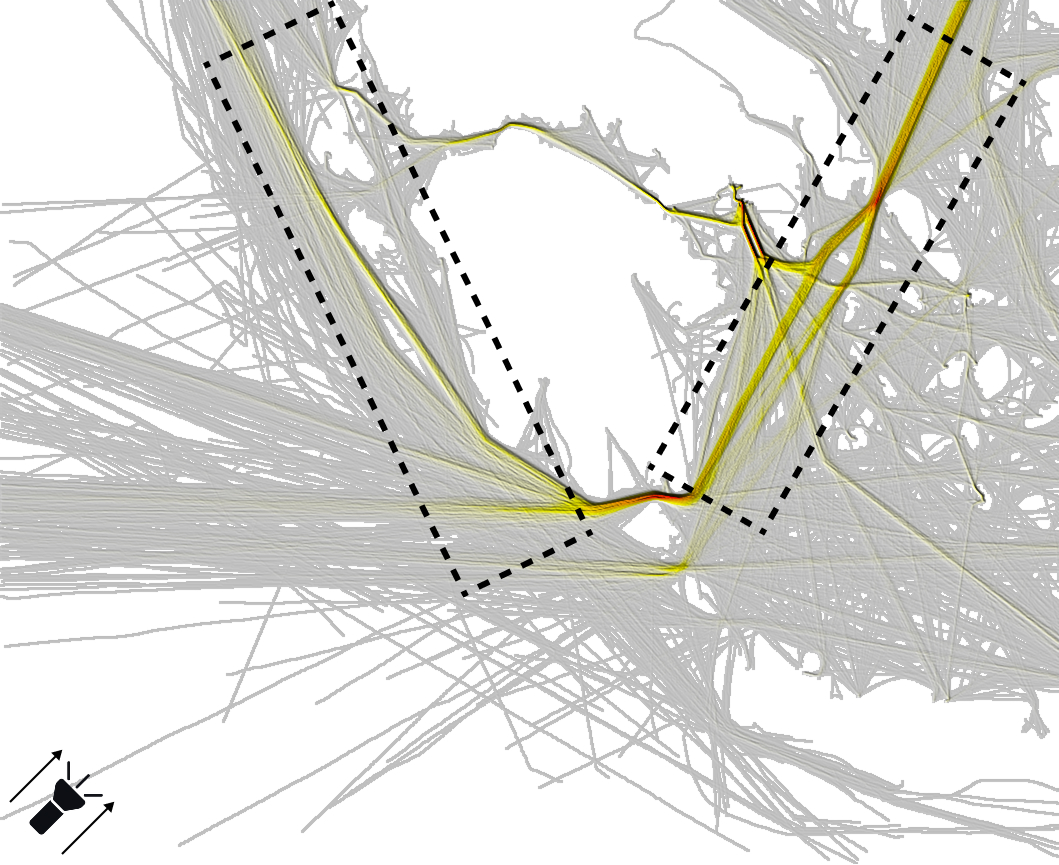}
        \caption{Lambertian shading with light from the southwest}
        \label{fig:left_bottom_light}
    \end{subfigure}
    \hfill
    \vspace{-4mm}
    \caption{The sensitivity of traditional Lambertian shading to a fixed global light direction. As the light comes from the northwest (left) versus the southwest (right), the set of highlighted line structures changes (see boxed section). %This demonstrates that a single global light direction fails to enhance structures with varying orientations uniformly.
    }
    \label{fig:different_light}
    \vspace{-5mm}
\end{figure}

Discretizing continuous trajectories into pixel-based bins can disrupt perceptual continuity, obscuring sequential connectivity and transitions between dense regions, as illustrated in \cref{fig:teaser:a}. This fragmentation complicates key tasks such as trend tracking and the detection of structural deviations in sparse areas. Existing illumination techniques for density plots, such as those by Willems~\textit{et al.}~\cite{willems2009vessel} and Scheepens~\textit{et al.}~\cite{scheepens2011composite}, often exacerbate these problems. A representative result is shown in \cref{fig:teaser:b}, where we isolate and apply only the illumination component of these methods. We omit their domain-specific preprocessing steps (e.g., filtering trajectories by vessel speed) to ensure a fair comparison, since our method is designed as a general line-based approach independent of application-specific attributes. By prioritizing density magnitude over flow, they introduce visual artifacts---including noise and color distortions---that obscure subtle line structures. Their reliance on kernel-size variation captures features only at a single scale while neglecting line orientation and shape, leading to weak representations of line structures in low-density regions.
Moreover, fixed global light directions do not account for local line alignments. As shown in \cref{fig:left_top_light,fig:left_bottom_light}, adjusting the light direction shifts the emphasis between different line sets (highlighted region), thereby substantially altering the visualization.
This approach is limited as its normals are tied only to the density, underrepresents low-density structures, and emphasizes lines based on light direction. These shortcomings reveal a mismatch between discrete pixel representations and the continuous nature of line data, motivating our new line shading strategy.

To address these challenges, we propose an illumination model designed to preserve continuity while selectively enhancing both dominant trends and sparse outliers. Our method consists of three key components, integrated into our pipeline (\cref{fig:pipeline}):

\noindent\textbf{Selective structure prioritization.}
Since line density plots typically visualize a large number of lines, it is infeasible to show all line details simultaneously. We therefore introduce a bin-based measure of \emph{outlierness}, which serves as a metric to quantify how much a line’s local tangent orientations deviate from those of its neighbors. A user-controlled parameter \textit{OutlierFocus} (\(\mu\)) determines which part of the outlierness spectrum is emphasized: when \(\mu=0\), trend-following trajectories dominate; when \(\mu=1\), strongly deviating outliers dominate; intermediate values highlight trajectories close to the chosen level of deviation. This user-tunable prioritization ensures that important structures are revealed without overwhelming the plot.
    
\noindent\textbf{Structural normal map construction.} The selective prioritization marks a key distinction from Lambertian methods, which derive normals only from density gradients. We regard these normals from Lambertian line shading methods~\cite{scheepens2011composite,scheepens2011interactive} as a low-frequency baseline \(\mathbf{n}_{\mathrm{low}}\) (\cref{fig:pipeline}b1). While this encodes main density trends, it cannot expose fine structural variation. We therefore construct a complementary high-frequency normal map \(\mathbf{n}_{\mathrm{high}}\) from trajectories prioritized by $\mu$ (\cref{fig:pipeline}b2). To combine these two normal maps into a final structural normal map \(\mathbf{n}_{\mathrm{structure}}\) (\cref{fig:pipeline}b3), we introduce the parameter \textit{StructureEmphasis} (\(\sigma\)) that acts as a selection threshold rather than a blending factor. It controls the proportion of top-ranked trajectories under a certain $\mu$ that contribute to the high-frequency map. This selection is applied on a per-line basis—a trajectory either contributes its full geometry or not at all, thus preserving its continuity. The final map is then composed using a prioritized replacement rule: for any given pixel, if a valid normal exists in the $\sigma$-filtered high-frequency map, it is used; otherwise, the map falls back to the low-frequency baseline.
%This composition scheme has intuitive results at its extremes.
When $\sigma=0$, the high-frequency map is empty, thus the composition rule defaults to the low-frequency baseline (\cref{fig:pipeline}b1) everywhere. When $\sigma=1$, all prioritized trajectories contribute, resulting in a structural map that layers the full high-frequency map (\cref{fig:pipeline}b2) over the low-frequency baseline. By adjusting $\sigma$, users can incrementally add or remove layers of fine-grained detail, smoothly transitioning between a pure trend-based visualization and one showing high-frequency structural information.
    
\noindent\textbf{Direction-adaptive illumination.}
To overcome the limitations of fixed global lighting of the Lambertian model~\cite{phongshading}, we estimate a dominant line orientation for each bin and set the local light vector perpendicular to this orientation, with an elevation angle of $60^\circ$ following perceptual recommendations~\cite{oshea2008assumed}. This yields an \emph{intensity map} (\cref{fig:pipeline}c) encoding locally adaptive Lambertian shading. Finally, following Chen~\textit{et al.}~\cite{chen2024visualization}, we apply the illumination exclusively to the luminance ($L$) channel of CIELAB while leaving chromatic channels ($a,b$) intact, thereby minimizing color distortion and preserving the density--color mapping.

The resulting illuminated density plot (\cref{fig:pipeline}d) maintains the readability of density while enhancing line continuity and selectively revealing structural outliers. Additional parameters, such as $\eta$ for normal-map scaling and $\phi$ for illumination blending (\cref{subsec:structure-enhancement}, \cref{subsection:colorcomposition}), further refine the rendering but are less critical for interactive use.

Together, these design choices target three goals: (\textbf{D1}) preserve and enhance perceived continuity in dense regions, (\textbf{D2}) enhance subtle and sparse structural deviations that density alone hides, and (\textbf{D3}) give users interactive control over the emphasis between trends and outliers.

\subsection{Line Outlierness}
\label{subsec:line-outlierness}

To distinguish between main trends and outliers, we propose a line outlierness measure based on bin-based line similarity. In our method, one bin corresponds to one pixel. Our approach adapts the distance field concept from scatterplot visualizations like Splatterplots~\cite{Mayorga2013Splatterplots}, where points diffuse influence to distinguish clusters from outliers. For lines, we repurpose the Continuous Density Estimation (CDE) field~\cite{lampe2011curve} of a line, treating its density distribution \(L_h(z)\) as a continuous \textit{influence field}. The similarity between two lines, \(l\) and \(l'\), is defined as the normalized integral of \(l'\) as it passes through the influence field of \(l\), modulated by their directional alignment:
\vspace{-2mm}
\begin{equation}
\text{sim}(l, l') = \frac{\oint_{l'} \left| \text{dir}(l, l', z) \right| \cdot L_h(z) \, dz}{\text{length}(l')}.
\label{eq:similarity}
\vspace{-2mm}
\end{equation}
Here, \(\text{dir}(l, l', z)\) is the dot product of the lines' direction vectors at point \(z\). We use its absolute value to capture similarity in orientation while disregarding the intrinsic directionality of the lines. This choice is crucial because line density plots are themselves direction-agnostic, aggregating paths based on their spatial congruence (i.e., shape) rather than their direction of travel.
This measure is notably non-commutative (\(\text{sim}(l, l') \neq \text{sim}(l', l)\)) because the integral path and the underlying influence field depend on the ordering of the lines.
This asymmetry is desirable, as it allows a long, dominant trend to exert a strong influence on a short line crossing it, while the short line has a small influence on the long trend.

\begin{figure}[tb]
    \centering
    \hfill
    \begin{subfigure}{0.23\linewidth}
        \centering
        \includegraphics[width=\linewidth]{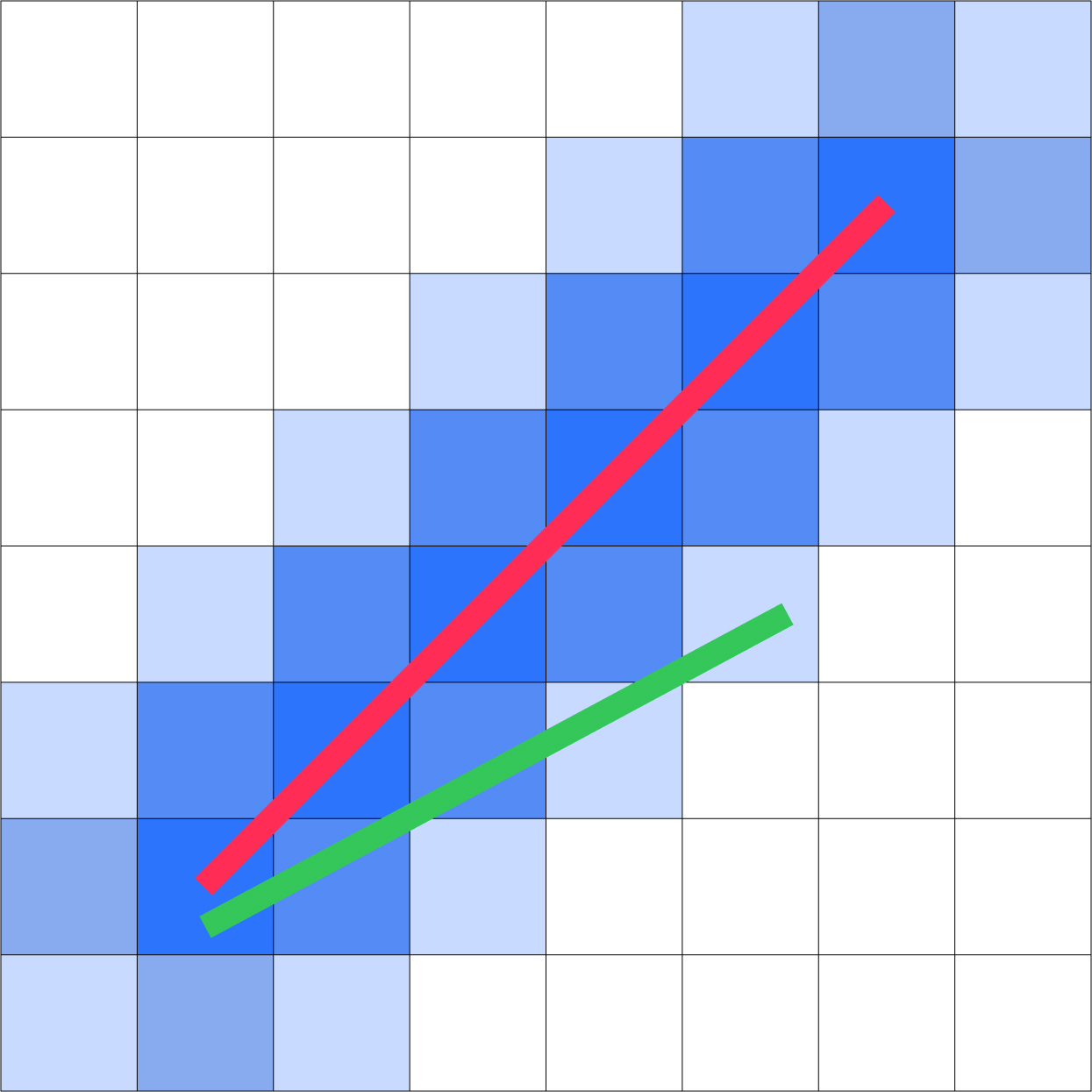}
        \caption{}
        \label{fig:line_similarity_a}
    \end{subfigure}
    \hfill
    \begin{subfigure}{0.23\linewidth}
        \centering
        \includegraphics[width=\linewidth]{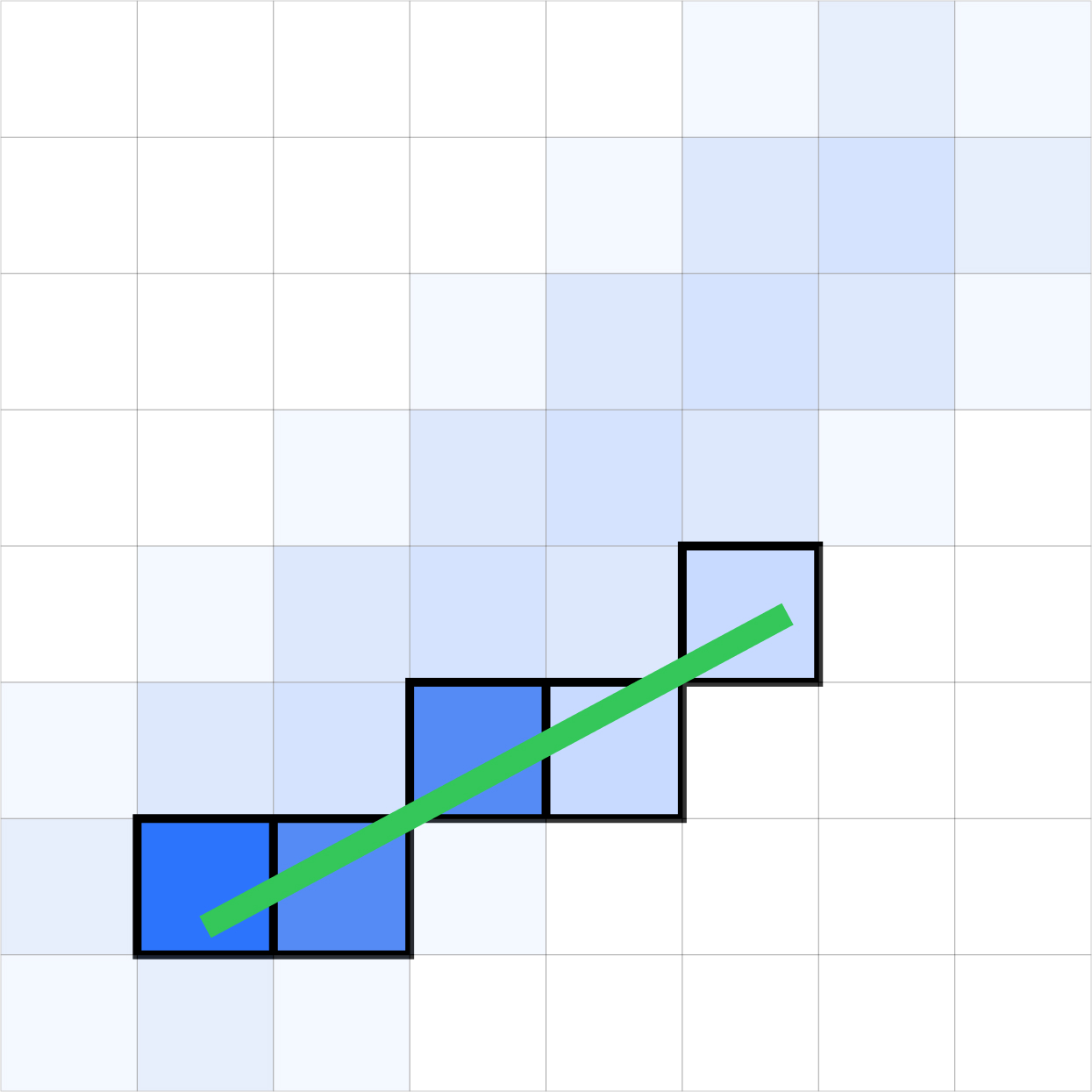}
        \caption{}
        \label{fig:line_similarity_b}
    \end{subfigure}
    \hfill
    \begin{subfigure}{0.23\linewidth}
        \centering
        \includegraphics[width=\linewidth]{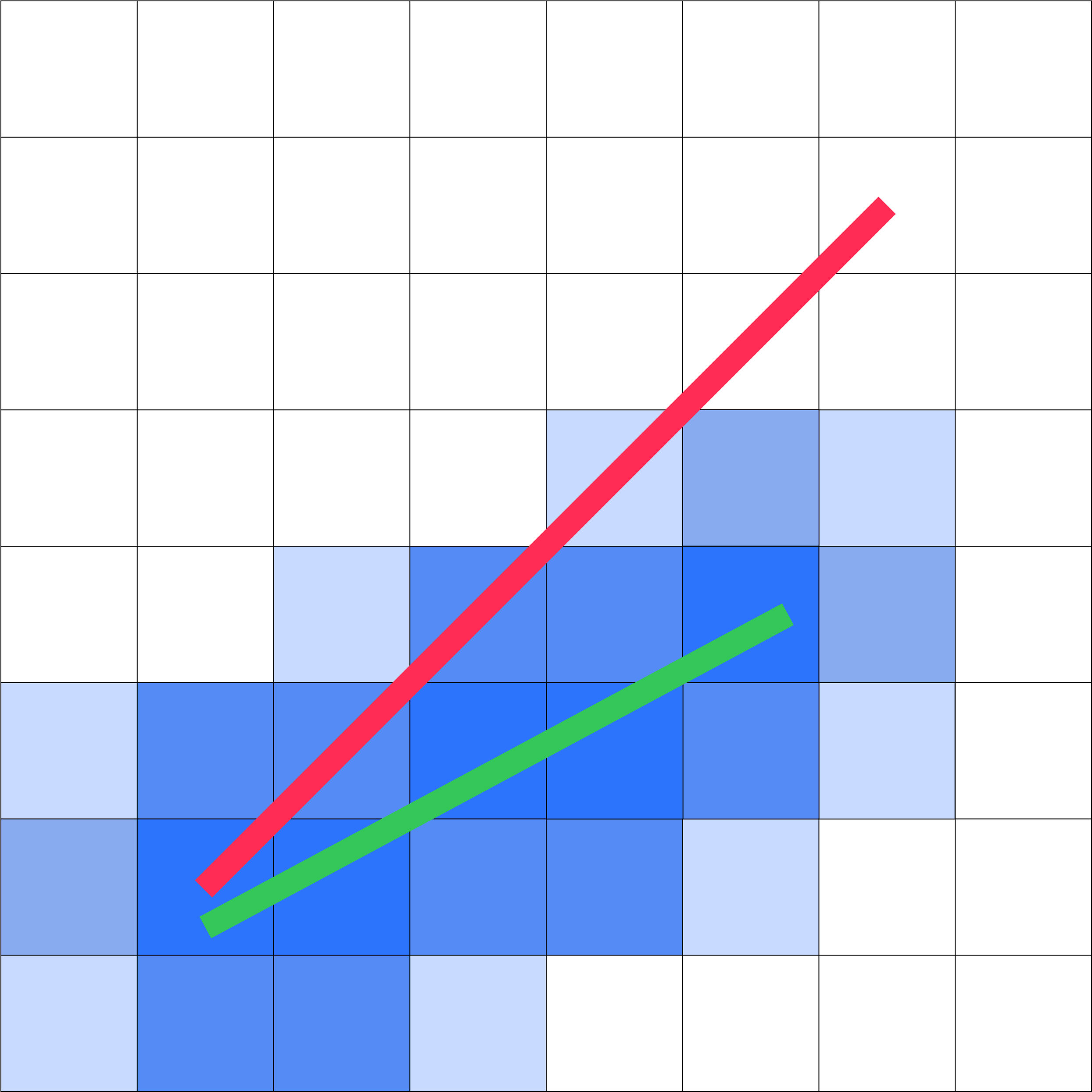}
        \caption{}
        \label{fig:line_similarity_c}
    \end{subfigure}
    \hfill
    \begin{subfigure}{0.23\linewidth}
        \centering
        \includegraphics[width=\linewidth]{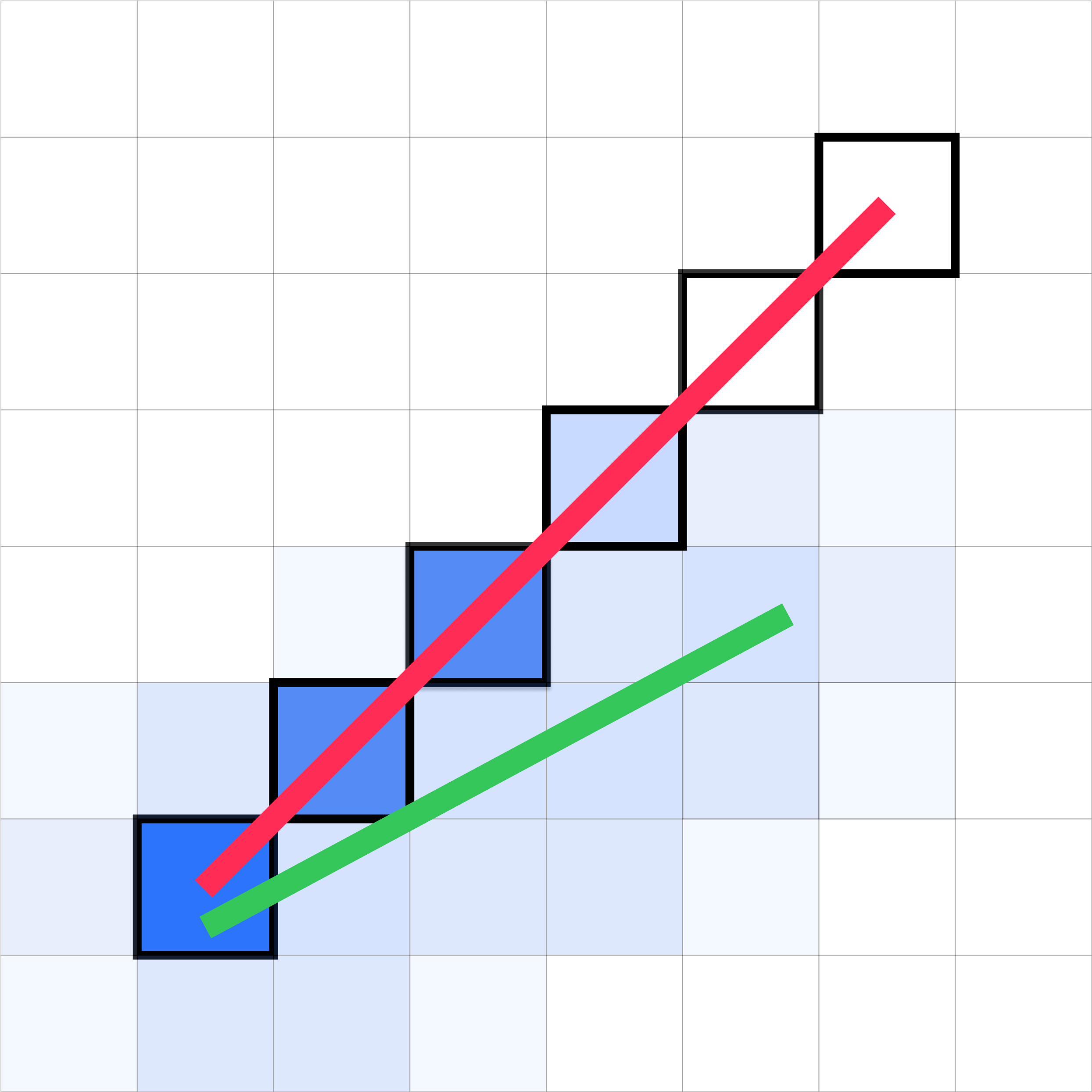}
        \caption{}
        \label{fig:line_similarity_d}
    \end{subfigure}
    \hfill
    \vspace{-3mm}
    \caption{Illustration of bin-based line similarity: (a) Construction of the distance field for a red line; (b) Discretized integral path over the red line’s field for a green line; (c) and (d) The discrete distance field of the green line, highlighting non-commutativity while limiting the diffusion range for efficiency.}
    \label{fig:line_similarity}
    \vspace{-5mm}
\end{figure}

To compute this efficiently, we discretize the integral. For a line \(l\), we precompute its influence field \(L_h\) at each grid bin, as illustrated in \cref{fig:line_similarity}. The similarity is then approximated by summing the influence values over the pixels, \(P_{l'}\), that constitute line \(l'\):
\vspace{-2mm}
\begin{equation}
\text{sim}(l, l') \approx \frac{\sum_{p \in P_{l'}} \left| \text{dir}(l, l', p) \right| \cdot L_h(p)}{|P_{l'}|}.
\label{eq:similarity_discrete}
\vspace{-2mm}
\end{equation}
To reduce computational costs, we only compute the influence field \(L_h\) within a narrow band around each line (e.g., diffusing influence perpendicularly up to 5 pixels on either side).

Using this similarity measure, we define a line’s outlierness as its average dissimilarity to its neighboring lines. First, for any given line \(l\), we define its set of neighboring lines, \(N(l)\). A line \(l' \neq l\) is considered a neighbor if at least one of its pixels lies within the pre-computed influence field of line \(l\). The outlierness of \(l\) is then formally defined as:
\vspace{-2mm}
\begin{equation}
\text{outlierness}(l) = 1 - \frac{1}{|N(l)|} \sum_{l' \in N(l)} \text{sim}(l, l')
\label{eq:outlierness}
\vspace{-2mm}
\end{equation}
where \(|N(l)|\) is the total number of unique neighboring lines. This formulation directly captures the average similarity of a line to its entire neighborhood. After computing the outlierness score for every line, we rank them from the lowest score (strongest inliers) to the highest (strongest outliers). This ranked list provides the crucial input for our user-driven structure emphasis model.

To validate our orientation-aware metric, we compared it against common geometric measures like Chamfer~\cite{butt1998optimum} and Hausdorff~\cite{Huttenlocher1993} distance. We found these standard metrics to be ill-suited for this task, as they are insensitive to local line shape and orientation, often incorrectly favoring short lines as inliers. Our orientation-aware metric is specifically designed to overcome these limitations. The detailed visual comparison and full analysis are provided in the supplementary material (see Sec.~2).

\subsection{User-Driven Line Structure Emphasis}
\label{subsec:structure-enhancement}

Our model decouples shading from the density colormap by constructing a \textbf{structural normal map} \(\mathbf{n}_{\mathrm{structure}}\) from two distinct sources: a \textbf{low-frequency normal map} \(\mathbf{n}_{\mathrm{low}}\) derived from the density field \(F(x,y)\)  as in prior work~\cite{scheepens2011interactive,scheepens2011composite}, and a \textbf{high-frequency normal map} \(\mathbf{n}_{\mathrm{high}}\) from individual lines. This process is guided by two user parameters: \textit{OutlierFocus} (\(\mu\)) and \textit{StructureEmphasis} (\(\sigma\)). The parameter \(\mu\) specifies the outlierness level to focus on, while \(\sigma\) controls the proportion of focused lines used to generate high-frequency details. For additional mathematical formulations, please refer to Sec.~1.1 of our supplementary material.

\noindent\textbf{Selective structure prioritization.}
We first normalize the outlierness ranks of all lines to a [0, 1] scale:
\vspace{-2mm}
\begin{equation}
l_i'=\frac{\mathrm{rank}(l_i)}{n-1},
\label{eq:outlierness_rank}
\vspace{-2mm}
\end{equation}
where \(l_i'=0\) is the strongest inlier. Based on the user's focus \(\mu\), we measure each line's proximity \(\delta_i(\mu)=|l_i'-\mu|\) and use \(\sigma\) to select a subset \(S_\sigma(\mu)\) of lines with the smallest \(\delta_i(\mu)\).

\noindent\textbf{High- and low-frequency normal computation.}
For each selected line, we define its spatial footprint \(B_i\) as the \(n\times n\) pixel window around the line, with the union of these footprints forming the high-frequency coverage set \(P_{\mathrm{high}}\). For a more detailed definition, refer to Sec 1.1 of our supplementary material.

The \textbf{low-frequency normal map} \(\mathbf{n}_{\mathrm{low}}\), which captures global density structure, is computed from the gradient of the density field \(F\):
\vspace{-1mm}
\begin{equation}
\mathbf n_{\mathrm{low}}(x,y)=
\frac{\bigl(-\partial_x F(x,y),\ -\partial_y F(x,y),\ 1/\eta\bigr)}
     {\sqrt{(\partial_x F(x,y))^2+(\partial_y F(x,y))^2+1/\eta^{2}}}.
\label{eq:low_freq_normal}
\vspace{-1mm}
\end{equation}
where the parameter \(\eta > 0\) controls the scaling of the normal's \(z\)-component. As \(\eta\) increases, the \(1/\eta\) term diminishes, causing the normal vector to be pulled toward the x-y plane. This alteration of the normal effectively exaggerates the perceived slope of the density field, thereby enhancing the visibility of subtle structures in areas with low gradients.
For the \textbf{high-frequency normal map}, we first define the influence field \(f_i(x,y)\) for each selected line, based on the CDE kernel \(L_h\) and perpendicular distance \(d_i(x,y)\):
\vspace{-1mm}
\begin{equation}
f_i(x,y)=L_h\bigl(d_i(x,y)\bigr), \quad \text{for } (x,y) \in B_i.
\label{eq:influence}
\vspace{-1mm}
\end{equation}
At pixels where footprints overlap, we select a single, unique contributing line \(c(x,y)\) to avoid averaging artifacts:
\vspace{-1mm}
\begin{equation}
c(x,y)=\arg\min_{i\in S_\sigma(\mu)\ \land\ (x,y)\in B_i}\bigl(\,\delta_i(\mu),\ d_i(x,y)\bigr).
\label{eq:contributor}
\vspace{-1mm}
\end{equation}
This selects the line closest to the focus \(\mu\), using distance \(d_i\) as a tie-breaker. The normal \(\mathbf{n}_{\mathrm{high}}\) is then derived from the gradient of this unique line's influence field, \(f_c\), which is detailed in Sec. 1.1 of our supplementary material.

\noindent\textbf{Final normal map composition.}
The final map is composed using a \textbf{prioritized replacement} rule, which we favor over alpha-blending to avoid generating misleading artifacts. This rule ensures every normal has a clear, interpretable origin:
\vspace{-2mm}
\begin{equation}
\mathbf n_{\mathrm{structure}}(x,y)=
\begin{cases}
\mathbf n_{\mathrm{high}}(x,y), & \text{if } (x,y)\in P_{\mathrm{high}},\\[4pt]
\mathbf n_{\mathrm{low}}(x,y),  & \text{otherwise}.
\end{cases}
\label{eq:structure_normal_map}
\end{equation}

\subsection{Line-aware Illumination}
\label{subsec:illumination}

With the structural normal map established, we now introduce a dynamic, line-aware illumination model to overcome the limitations of traditional fixed lighting. Our approach computes a light direction \(\mathbf{l}(x, y)\) for each pixel, optimized to reveal local line structures. The process involves two main stages: first, computing the local light direction based on line orientation; second, shading the structural normal map \(\mathbf{n}_{\mathrm{structure}}\) with this light to produce a final intensity map \(I(x, y)\).  For detailed mathematical definitions, please refer to Sec.~1.2 of our supplementary material.

\noindent\textbf{Computing local light direction.}
The core principle of our model is that light should strike lines perpendicularly to maximize the visibility of their shape and detail. Finding the dominant orientation \(\mathbf{d}(x, y)\) differs for high- and low-frequency regions:

\begin{itemize}
    \item \textbf{For high-frequency pixels (\((x,y) \in P_{\mathrm{high}}\))}, the goal is to illuminate the specific, user-prioritized line structure. The dominant direction \(\mathbf{d}(x, y)\) is therefore directly set to the direction vector of the unique contributing line \(c(x,y)\) (from Eq.~\eqref{eq:contributor}) at that pixel.

    \item \textbf{For low-frequency pixels (\((x,y) \notin P_{\mathrm{high}}\))}, we perform a weighted Principal Component Analysis (wPCA)~\cite{costa09weighted} on the local direction vectors (tangents) of all trajectories in the pixel's neighborhood (detailed definition see Sec. 1.2 of supplementary material). The principal eigenvector from the wPCA defines the axis of dominant orientation. To resolve its inherent sign ambiguity and ensure a stable direction, its orientation is deterministically set by aligning it with the weighted mean direction of the same local tangents.
\end{itemize}

Once the dominant 2D orientation \(\mathbf{d}(x, y)\) is determined, the light's horizontal projection is set perpendicular to it. This 2D vector is then lifted into 3D with a fixed \(60^\circ\) elevation angle~\cite{oshea2008assumed} and normalized to form the final light vector \(\mathbf{l}(x, y)\). The detailed formulas for this process and the wPCA are in the supplementary material.

\noindent\textbf{Computing the intensity map.}
With both the structural normals and local light vectors defined, we compute the illumination intensity using the Lambertian model~\cite{phongshading} only, which avoids specular highlights that can distort the density colormap and distract from the visualization focus~\cite{chen2024visualization, rusinkiewicz06exaggerated}:
\vspace{-1mm}
\begin{equation}
I(x, y) = \mathbf{n}_{\mathrm{structure}}(x, y) \cdot \mathbf{l}(x, y).
\label{eq:diffuse_intensity}
\end{equation}
The resulting \textit{intensity map} \(I(x, y)\), shown in \cref{fig:pipeline}(c), is then ready for the final color composition stage.

\subsection{Color Composition}
\label{subsection:colorcomposition}
To integrate the intensity map into the original density plot without compromising perceptual encoding, we combine them in a way that minimizes distortions to hue and saturation (e.g., the A/B channels in CIELAB space). Traditional illumination models, such as Phong reflectance, might alter these components, producing visual artifacts like hue shifts (e.g., blue high-density regions turning greenish under specular highlights) or saturation over-boosting, which obscure density gradients and hinder trend tracking (D1) or outlier detection (D2).

Following Chen \textit{et al.}~\cite{chen2024visualization}, we apply the illumination exclusively to the luminance channel, ensuring that hue and saturation remain unchanged. First, the intensity map \( I(x, y) \) is scaled to normalize its range and preserve empty regions:
\vspace{-1mm}
\begin{equation}
I'(x, y) = \phi \cdot \frac{I_{\text{empty}} - I(x, y)}{I_{\text{empty}} - I_{\min}},
\label{eq:scaled_intensity}
\vspace{-1mm}
\end{equation}
where \( \phi \) controls the magnitude of the shading (as detailed in \cite{chen2024visualization}), \( I_{\text{empty}} \) is the intensity for empty areas, and \( I_{\min} \) is the minimum intensity.

For multi-hue colormaps used by Xue~\textit{et al.}~\cite{xue2024reducing}, we convert the density plot from an RGB color space to LAB, add \( I'(x, y) \) to the L channel, and clamp it to [0, 100] before converting back to RGB. For single-hue colormaps, a similar adjustment is applied in HCL space to maintain constant hue. This additive approach avoids disproportionate effects on high-luminance regions, supporting structural enhancement of the high/low-frequency normal maps without compromising density perception. Results for both colormap types are discussed in \cref{sec:evaluation}.

\begin{figure*}[tb]
    \centering
    \begin{subfigure}{0.24\textwidth}
        \centering
        \includegraphics[width=\textwidth]{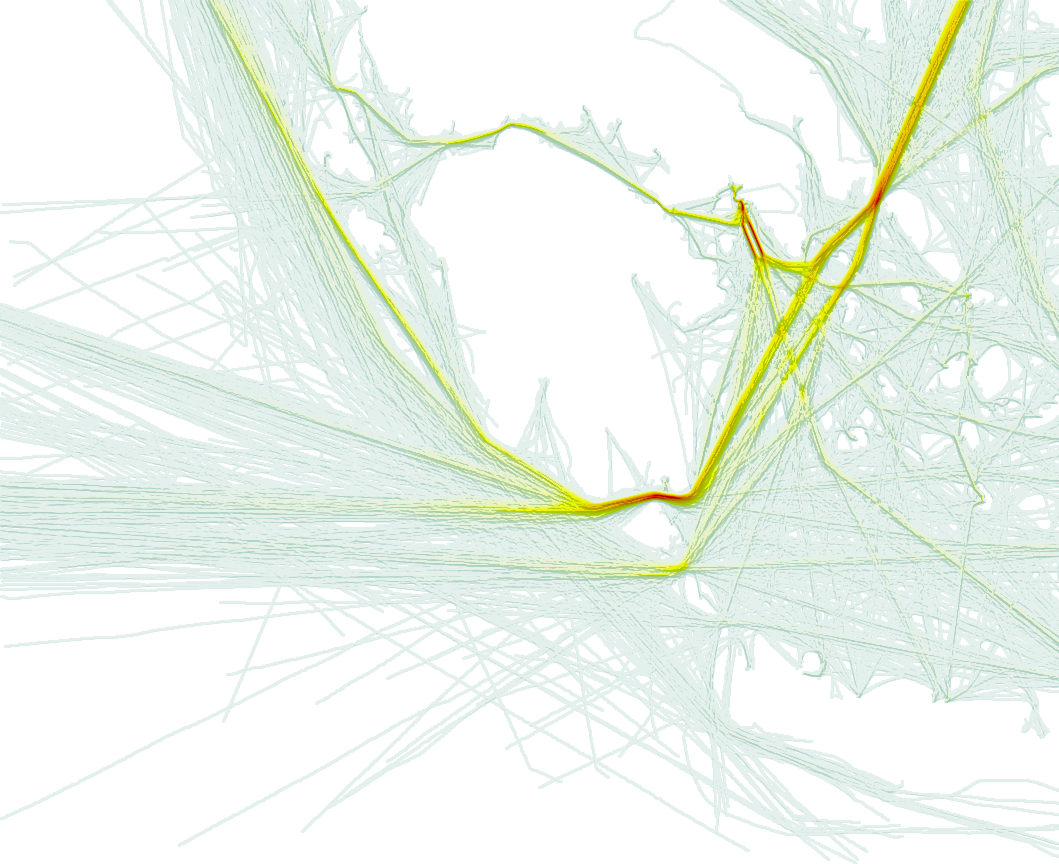}
        \caption{Without structure emphasis}
        \label{fig:ship_without_normal}
    \end{subfigure}
    \begin{subfigure}{0.24\textwidth}
        \centering
        \includegraphics[width=\textwidth]{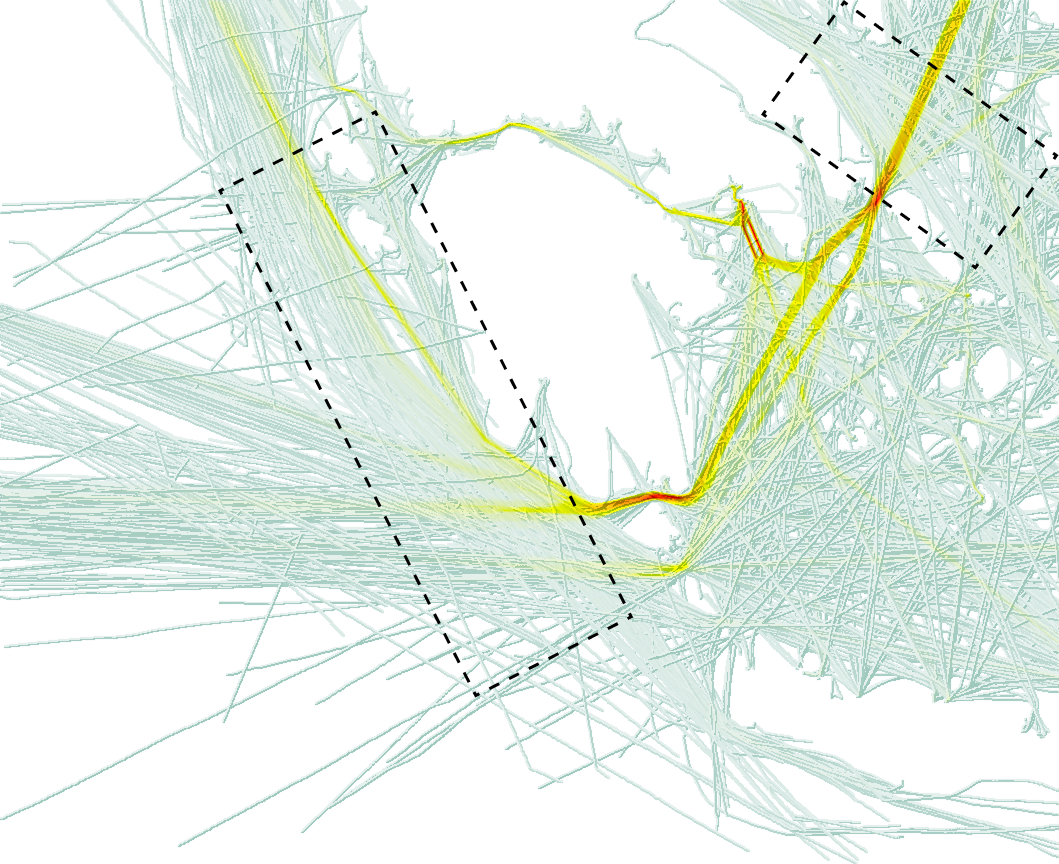}
        \caption{Without bin-based light computation}
        \label{fig:ship_without_light}
    \end{subfigure}
    \begin{subfigure}{0.24\textwidth}
        \centering
        \includegraphics[width=\textwidth]{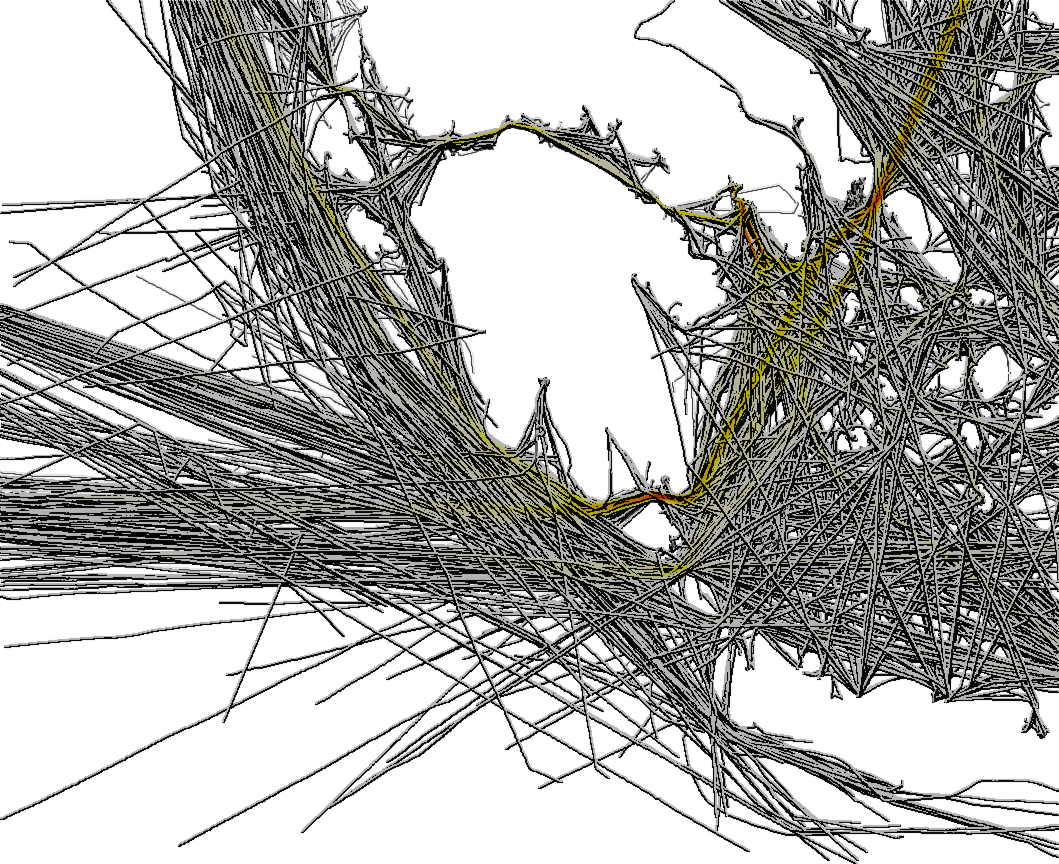}
        \caption{Without color composition}
        \label{fig:ship_without_color}
    \end{subfigure}
    \begin{subfigure}{0.24\textwidth}
        \centering
        \includegraphics[width=\textwidth]{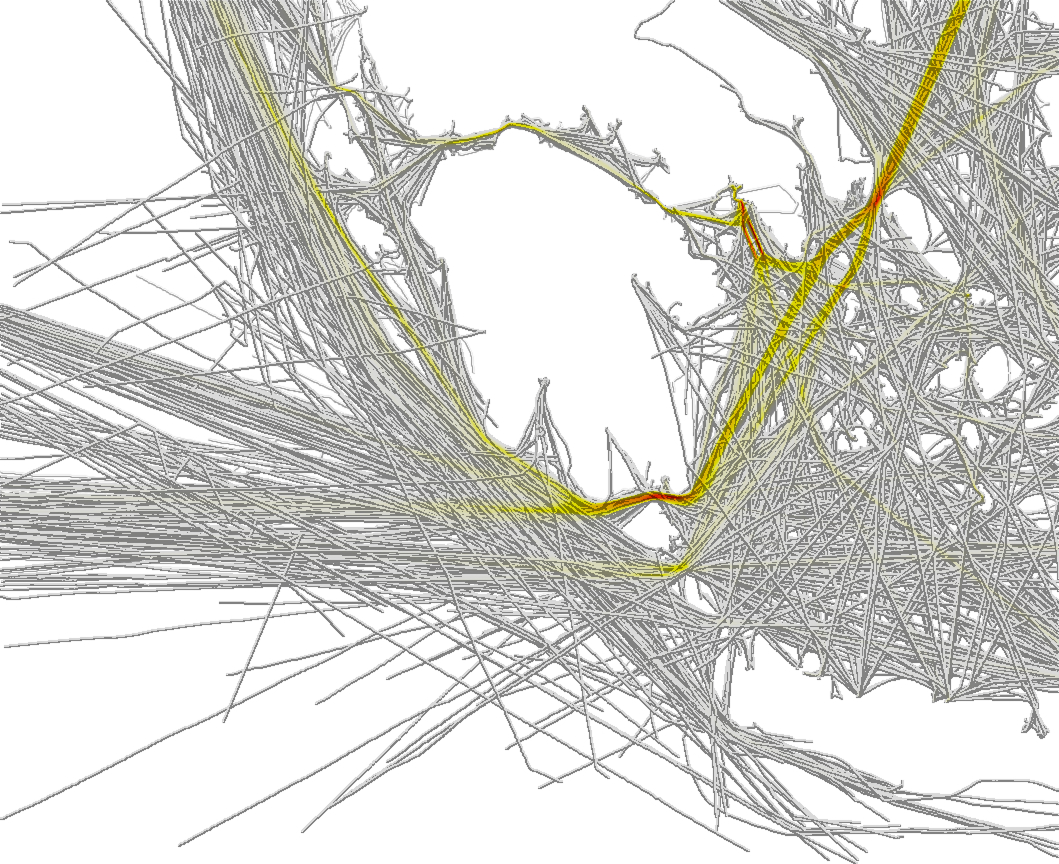}
        \caption{Without color composition (scaled darkening)}
        \label{fig:ship_without_color_scale}
    \end{subfigure}
    \vspace{-3mm}
    \caption{
    Ablation and comparative study on the vessel dataset~\cite{frantzis2018hellenic} (cf. \cref{fig:teaser}): (a) Without structure emphasis, the image appears flat and lacks detail; (b) A fixed global light introduces strong orientation bias (dashed box); (c) Direct RGB shading causes severe color distortion; (d) Scaled RGB shading still produces significant color distortion.
    }
    \label{fig:ablation}
    \vspace{-5mm}
\end{figure*}

\subsection{Interaction}
\label{subsec:interaction}
Based on the algorithmic components defined in this section, we implemented an interactive visualization system that enables users to adjust model parameters (\cref{fig:pipeline}) and explore visualization results in real time. As demonstrated in the accompanying video, our implemented system interface comprises three primary components:
% Leveraging the computational efficiency of modern graphics hardware, our bin-based shading method is optimized for real-time interactive manipulations. Our interactive system comprises three primary components (as shown in the accompanying video):

\noindent\textbf{Density plot settings.} Users can customize the color mapping scheme of a density plot, selecting either a single-hue or multi-hue colormap. They are able to define the number of clusters, perform the clustering computation, and assign hues to each cluster, adhering to the approach outlined by Xue {\textit{et al.}}~\cite{xue2024reducing}.

\noindent\textbf{Parameter settings.} The system incorporates interactive sliders, enabling users to adjust parameters of our model in real time. These include the \textit{OutlierFocus} ($\mu$) and \textit{StructureEmphasis} ($\sigma$) (see \cref{subsec:structure-enhancement}), the deviation factor \(\eta\) (refer to \cref{eq:low_freq_normal}), and the scaling factor \(\phi\) (as defined in \cref{eq:scaled_intensity}).

\noindent\textbf{Light adjustments.} In addition to the automated lighting optimization mechanism, our system supports manual adjustments of light directions to accommodate user preferences. Manual light interaction is implemented on a per-cluster basis, with each cluster associated with a dedicated interaction point. Users can drag and drop this point to modify the azimuth and elevation of the light source for that respective cluster. Each point features a range control mechanism, which defines a sector of permissible light azimuths surrounding it. 
\section{Evaluation and Results}
\label{sec:evaluation}

In this section, we first rigorously evaluate the core components of our model and then present our results on real-world data. This evaluation proceeds in four stages: (i) an ablation study isolating user-driven structure emphasis, bin-based light-direction computation, and luminance-only composition; (ii) a quantitative color-fidelity analysis using the CIEDE2000 metric on three datasets; (iii) quantitative performance and scalability evaluation of our pipeline's components; and (iv) a set of qualitative case studies comparing our method against plain density plots and diffuse shading in both single- and multi-cluster settings (adapting the work of Xue~\textit{et al.}~\cite{xue2024reducing}).

\subsection{Ablation and Comparative Evaluation}
\label{subsec:ablation_study}
We performed several ablation and comparative studies on the vessel dataset to assess the contribution of each component in our method. Variants with omitted components and alternative techniques, such as a 2D transfer function, were compared against the full approach (see \cref{fig:ship_1.0_0.5_3.0_-20}) under the parameter configuration $\mu=1.0$, $\sigma=0.5$, $\eta=3.0$, $\phi=-20.0$, and kernel size $3\times3$.
The following comparisons are based on qualitative visual assessment by the authors, not on formal perceptual metrics or a controlled user study. Our assessment is grounded in three specific visual criteria relevant to line-data analysis and our design goals: (\textbf{C1}) structural continuity, the ability to perceive and follow individual line trends, especially through dense or intersecting regions; (\textbf{C2}) outlier visibility, the perceptibility of sparse trajectories that deviate from dominant trends; and (\textbf{C3}) color fidelity, the degree to which the original density-to-color mapping is preserved after illumination.

\noindent\textbf{Impact of removing user-driven line structure emphasis.} \cref{fig:ship_without_normal} shows the result without our user-driven line structure enhancement (\cref{subsec:structure-enhancement}). In this configuration, the pipeline defaults to using only the low-frequency normal map, which is derived directly from the density gradient. This approach fundamentally fails to decouple structural normals from the density map, as the normals become a direct function of density. Consequently, it performs poorly against our evaluation criteria: (1) While it allows us to partially highlight the structural continuity (\textbf{C1}) of major, high-density patterns, it (2) completely fails to enhance outliers (\textbf{C2}), as sparse trajectories do not generate a significant gradient to contribute to the normal map. The resulting flat appearance, which lacks structural separation, underscores the critical role of our high-frequency, trajectory-aware normals.

\noindent\textbf{Impact of removing bin-based light computation.} In \cref{fig:ship_without_light}, we omit the bin-based light computation (\cref{subsec:illumination}), using a fixed light source at the top-left corner instead. This introduces a strong orientation bias (see dashed box), which directly compromises our first two evaluation criteria: depending on a line's orientation relative to the light, both \textbf{C1} for trends and \textbf{C2} for sparse lines can be significantly compromised. This demonstrates that our bin-based, adaptive light direction is essential for robustly revealing structures and outliers regardless of their orientation.

\noindent\textbf{Impact of removing color composition.} 
Omitting the color composition step (\cref{subsection:colorcomposition}) and shading directly in RGB channels (\cref{fig:ship_without_color}) introduces color shifts (i.e., hue and saturation changes) that are prominent when compared to the original density map. Scaling the darkening effect (\cref{fig:ship_without_color_scale}) still alters colors substantially. These severe distortions obscure the density field, impairing density interpretation. This demonstrates a clear failure of criterion \textbf{C3}, underscoring that CIELAB-based, luminance-only composition is essential for preserving the perceptual accuracy of the density map.

We further compared our model against image-space techniques: histogram equalization (\texttt{histeq}), adaptive histogram equalization (\texttt{adapthisteq}), and a 2D transfer function (visual results and analysis in Sec.~3 of supplementary material). In summary, our integrated components---user-driven line structure emphasis, bin-based light adjustment, and color composition---collectively ensure detailed structural rendering, adaptive illumination, and accurate density representation in complex line-based visualizations.
% To further validate our illumination model, we compared it against other image-space enhancement techniques, namely histogram equalization (histeq), adaptive histogram equalization (adapthisteq), and a 2D transfer function. We provide visual results and analysis for all these comparisons in  Sec.~3 of our supplementary material.
% In summary, the integrated components of our method---user-driven line structure emphasis, bin-based light adjustment,
% and color composition---collectively ensure detailed structural rendering, adaptive illumination, and accurate density representation in complex line-based visualizations.

\begin{figure}[tb]
    \centering
    \includegraphics[width=0.95\linewidth]{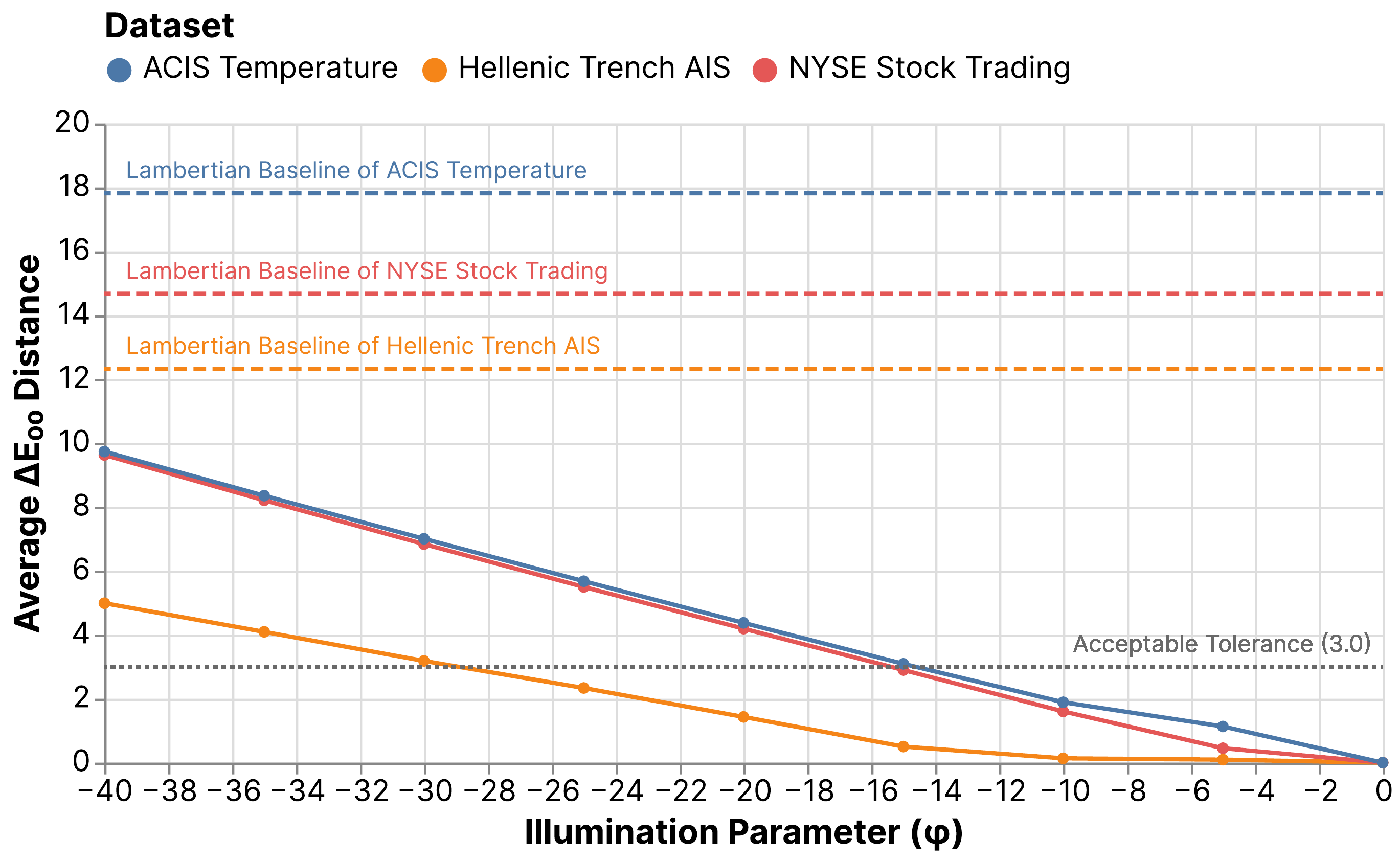}
    \vspace{-3mm}
    \caption{
   Color difference ($\Delta E_{00}$) as a function of the illumination parameter $\phi$. Solid lines trace our method's distortion per dataset, while dashed lines show the higher, fixed distortion of the Lambertian baseline. The gray line marks our acceptable tolerance threshold of $\Delta E_{00} = 3.0$, based on printing industry benchmarks~\cite{liu13discussion}.
    }
    \label{fig:phi_vs_ciede}
    \vspace{-6mm}
\end{figure}

\subsection{Color Distortions}
\label{subsec:color_distortions}

Our illumination model enhances structural visibility by modulating the luminance channel, a process controlled by the scaling factor $\phi$. A key consideration is the trade-off between the strength of this enhancement and the resulting color distortion. When $\phi = 0$, no illumination is applied, and the color distortion is zero. As $\phi$ becomes more negative, the illumination effect is amplified, increasing the perceptual distance from the original density colormap.
To systematically analyze this trade-off, we measured the average per-pixel CIEDE2000 color difference ($\Delta E_{00}$)~\cite{sharma2005ciede2000} while varying the $\phi$ parameter from 0 to -40. The experiment was conducted on the three datasets introduced in \cref{subsec:case_study} using a multi-hue colormap. For this calculation, we only considered non-empty pixels where at least one line contributes to the density. The results are presented in \cref{fig:phi_vs_ciede}.

This analysis provides a practical guide for parameter selection. Although formal perceptual tolerances for density plot visualizations are lacking, threshold values can be informed by domains where color fidelity is essential. A study by Liu~\textit{et al.} on printing industry tolerances found that a previously accepted ISO standard corresponded to an average CIEDE2000 difference of $\Delta E_{00} = 3.3$~\cite{liu13discussion}. For a broader context, even for the older CIELAB formula, a $\Delta E_{76}$ value above 3.5 is considered a clearly noticeable difference~\cite{mokrzycki11color}. Grounded in these practical studies, we adopt a conservative threshold of $\Delta E_{00} = 3.0$ to represent an acceptable perceptual difference. This allows an analyst to balance structural enhancement with color fidelity.

The results in \cref{fig:phi_vs_ciede} lead to two conclusions. First, our method consistently and significantly outperforms the Lambertian baseline. This significant difference is by design. The baseline's high distortion (see dashed lines) is a direct consequence of applying Lambertian shading---which acts as a form of brightness scaling---directly in the RGB color space. This single operation non-linearly alters all three dimensions of the perceptual CIELAB space (L, a, and b), causing visible shifts in both hue and luminance. Our method, by contrast, is designed to apply illumination only to the L channel, which results in fundamentally lower color distortion. Across all tested datasets, our method's color distortion remains far below that of traditional Lambertian shading. Second, the chart quantifies the data-dependent cost of enhancement. For instance, the Hellenic Trench AIS dataset exhibits a gradual trade-off, where a strong illumination parameter of $\phi = -25$ yields a significant visual enhancement while maintaining an average color distortion of $\Delta E_{00} \approx 2.3$, which is well within our acceptable tolerance of 3.0. In contrast, the NYSE Stock Trading and ACIS Temperature datasets exhibit a stronger distortion, crossing the acceptable tolerance threshold around $\phi \approx -15$.
This highlights that the perceptual distortion of our technique is data-dependent, providing an explicit guide for analysts to make task-specific choices. For an analysis focused on the precise interpretation of density values from the colormap, one might select a conservative parameter (e.g., $\phi=-15$ for the stock data) to ensure color fidelity. However, for an exploratory task where the primary goal is to reveal the faintest structural details and outliers, an analyst might intentionally choose a stronger illumination effect that exceeds the 3.0 tolerance, accepting a more noticeable color shift as a reasonable trade-off for maximal structural clarity.

\subsection{Quantitative Performance Evaluation}
We evaluate the computational performance of our line-based visualization technique, focusing on three core components: outlierness (measuring line deviation from local trends), normal map generation, and line-aware lighting. Outlierness is computed once at the start, as it remains static during interaction. However, it has the longest runtime due to the pairwise similarity calculations. In contrast, normal map and lighting computations are updated with each interaction, making their efficiency critical for real-time performance. Experiments were conducted on a laptop with Apple Silicon M1, 16GB memory, and Google Chrome (version 130, current at testing). We used JavaScript’s \texttt{console.time()} function to measure the runtime and averaged over 10 runs to reduce variability. As a dataset, we utilized real-world vessel trajectories from Hellenic Trench AIS data, with a fixed image resolution of 1059 x 864 pixels (matching the geographic data ratio) to match typical display ratios.

\begin{figure}[t]
\centering
\includegraphics[width=0.9\linewidth]{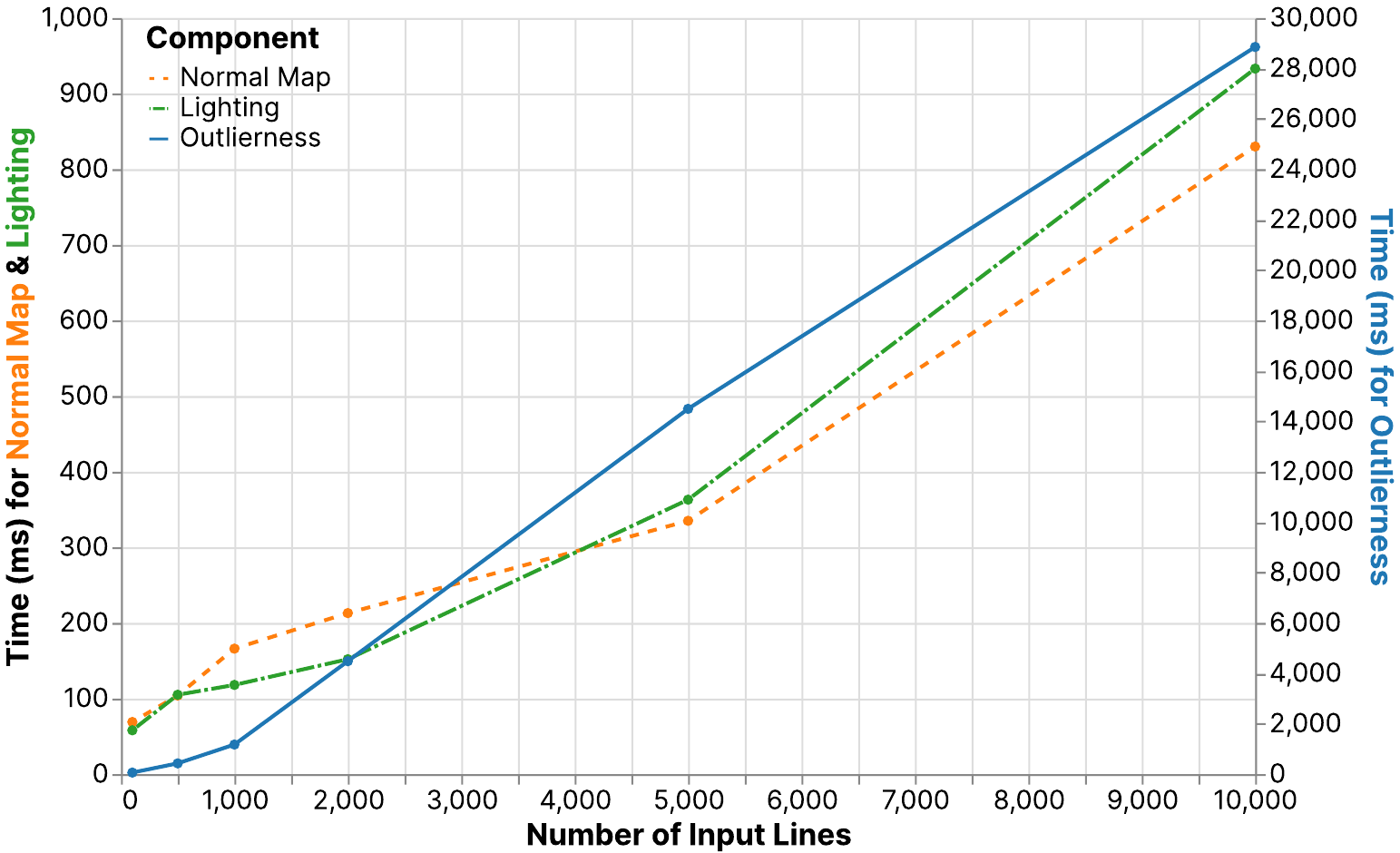}
\vspace{-3mm}
\caption{Scalability of computation time vs. number of input lines for vessel trajectory visualization on an Apple M1 laptop. Solid blue line: outlierness; dashed orange line: normal map; dot-dashed green line: lighting.}
\label{fig:scalability}
\vspace{-6mm}
\end{figure}

To assess scalability, we randomly sub-sampled 100 to 10,000 (100, 500, 1000, 2000, 5000, and 10000) lines from the original dataset, reflecting varying dataset sizes. Outlierness computation (\cref{subsec:line-outlierness}) involves pixel-based similarity integrals with a 5-pixel diffusion range. Since our method calculates the average integral sum of one line with all other lines, we achieve near-linear scaling ($O(n)$) by precomputing the weighted summed vector information from all lines in each bin, avoiding $O(n^2)$ complexity for pairwise comparisons, but with significant runtime (e.g., ~28.8 seconds for 10,000 lines, see \cref{fig:scalability}). Normal map computation (\cref{subsec:structure-enhancement}) involves generating a high-frequency map and a structure map (blended with a low-frequency map), while lighting computation (\cref{subsec:illumination}) utilizes a 3x3-pixel kernel and depends on parameters such as $\sigma$ (proportion of feature selection). 
As shown in \cref{fig:scalability}, which uses two y-axes, normal map and lighting exhibit linear growth ($O(n)$), since the wPCA on 2D vectors has a closed-form solution and its complexity is thus dominated by the single pass to build the covariance matrix.
Both components have update times of less than 1 second (left y-axis) for up to 10,000 lines, while outlierness is plotted on the right y-axis to accommodate its larger range. 
Because an interactive update requires recalculating both, the total update delay can approach 2 seconds.
% As shown in \cref{fig:scalability}, which uses two y-axes, normal map and lighting exhibit linear growth ($O(n)$) with times less than 1 second (left y-axis) for up to 10,000 lines, enabling interactive updates, while outlierness is plotted on the right y-axis to accommodate its larger range. %The dual linear y-axes clearly display the linear trends of all components as straight lines.
Both the light and normal map generation, as well as the outlierness computation, are linear, but with different time scales.
These results confirm that outlierness dominates total computation time but is performed once, while normal map and lighting support real-time interaction for medium-scale datasets (up to 10,000 lines), achieving responsive visualization. Bottlenecks in outlierness calculations for large $n$ could be optimized using GPU acceleration or approximate nearest-neighbor methods (e.g., k-d trees).

\label{subsec:performance}

\begin{figure*}[tbp]
    \centering
       \begin{subfigure}[b]{0.33\textwidth}
        \centering
        \includegraphics[width=\textwidth]{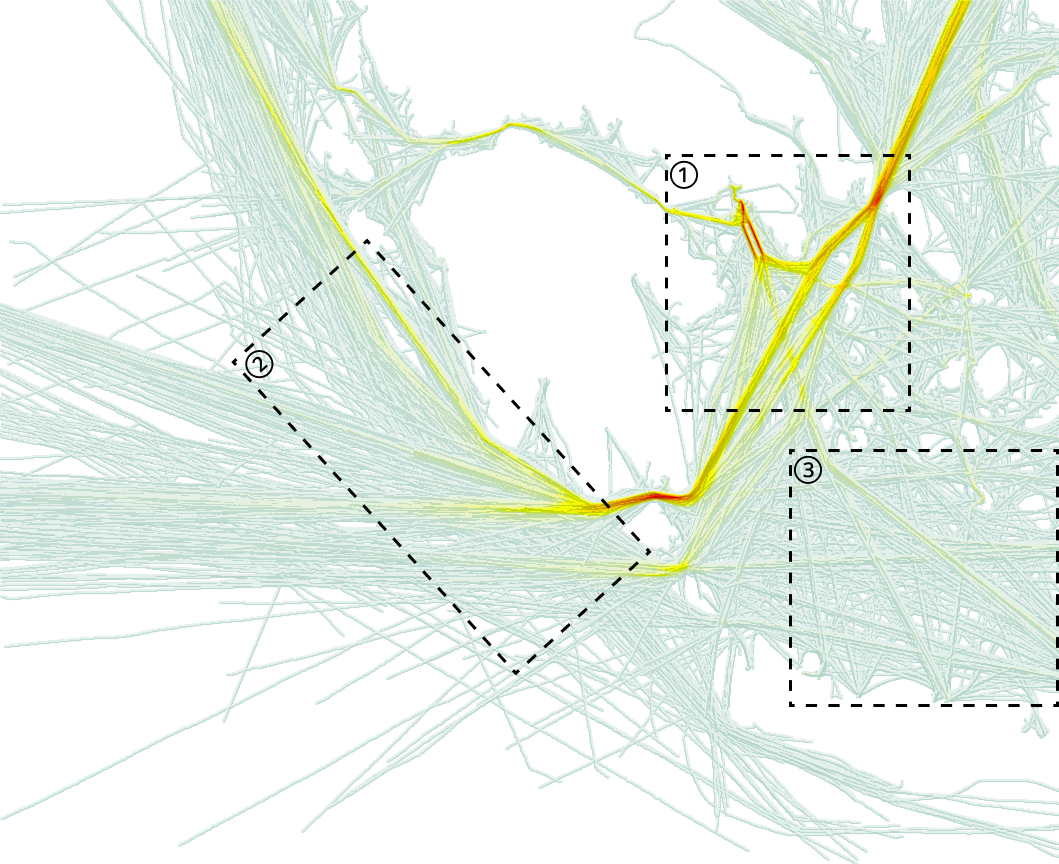}
        \caption{Our method when $\mu=0.0$, $\sigma=0.5$, $\eta=3.0$, $\phi=-20.0$}
        \label{fig:ship_0.0_0.5_3.0_-20}
    \end{subfigure}
    \hfill
    \begin{subfigure}[b]{0.33\textwidth}
        \centering
        \includegraphics[width=\textwidth]{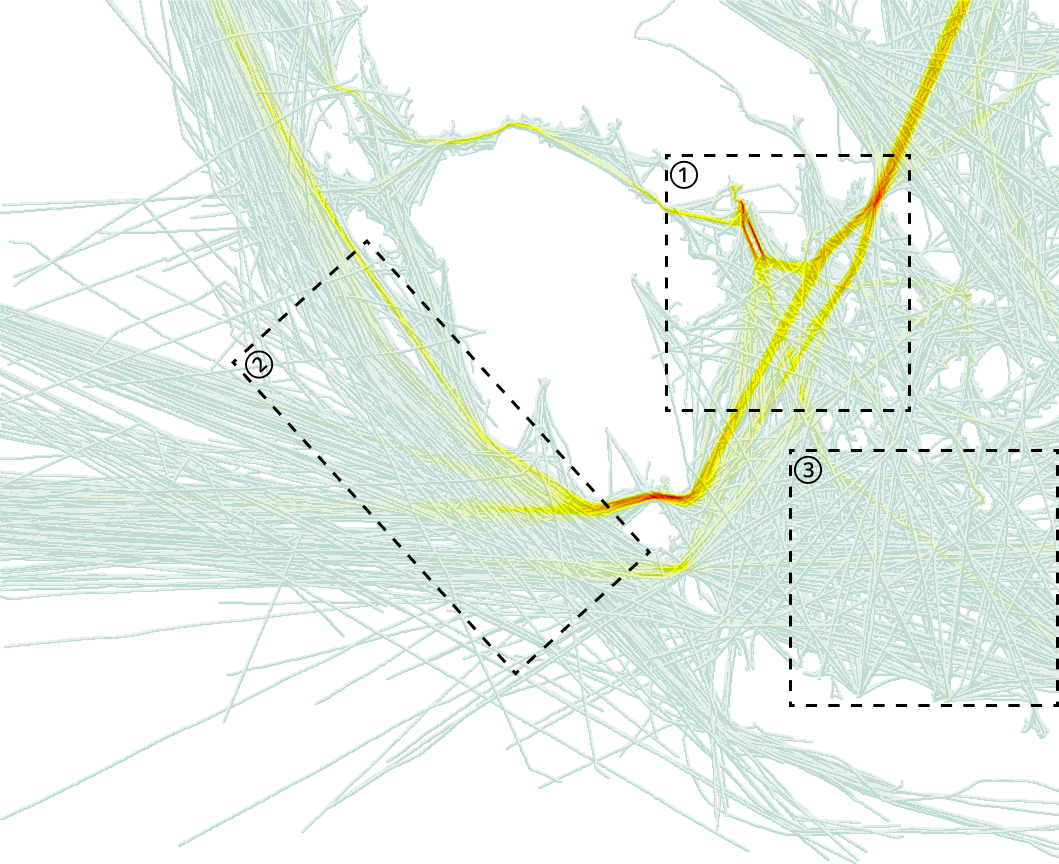}
        \caption{Our method when $\mu=1.0$, $\sigma=0.5$, $\eta=3.0$, $\phi=-20.0$}
        \label{fig:ship_1.0_0.5_3.0_-20}
    \end{subfigure}
    \hfill
    \begin{subfigure}[b]{0.33\textwidth}
        \centering
        \includegraphics[width=\textwidth]{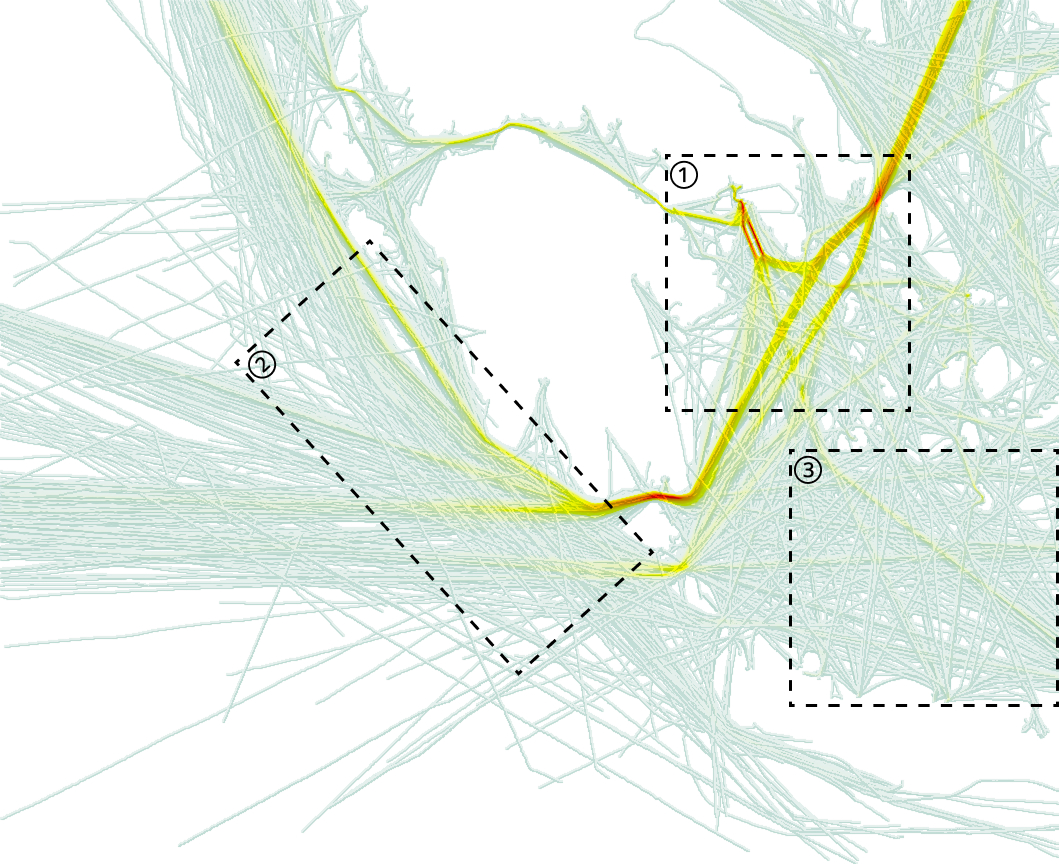}
        \caption{Our method when $\mu=1.0$, $\sigma=0.05$, $\eta=3.0$, $\phi=-20.0$}
        \label{fig:ship_1.0_0.05_3.0_-20}
    \end{subfigure}
    \hfill
    \vspace{-6mm}
    \caption{Our method with different parameters to highlight different degrees of outlierness for the Hellenic Trench AIS dataset~\cite{frantzis2018hellenic}. A $3\times3$ kernel generates all of the results.}
    \label{fig:ship_multihue}
    \vspace{-1mm}
\end{figure*}

\begin{figure*}[tbp]
    \centering
    \begin{subfigure}[b]{0.33\textwidth}
        \centering
        \includegraphics[width=\textwidth]{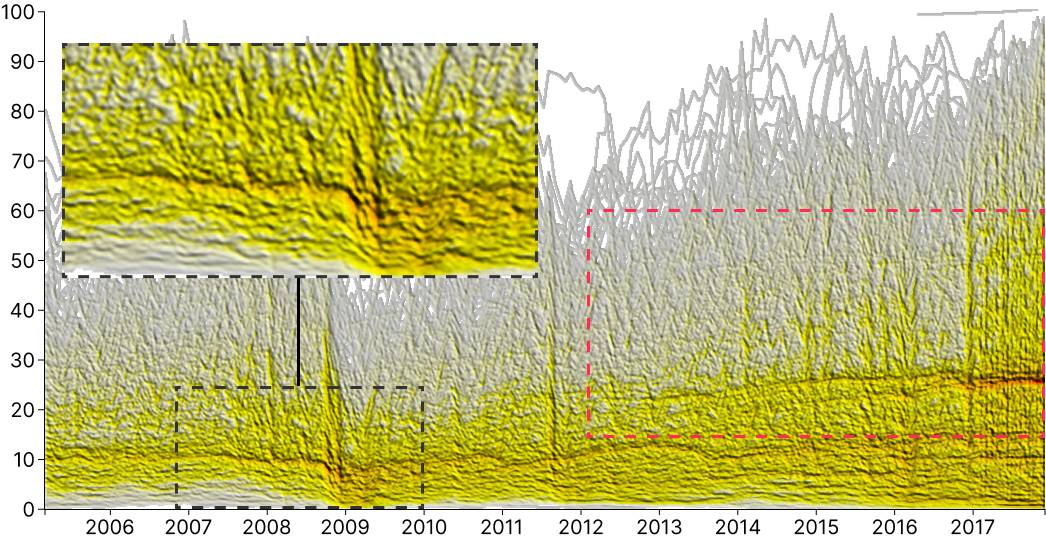}
        \caption{Direct Lambertian model}
        \label{fig:stock:a}
    \end{subfigure}
    \hfill
    \begin{subfigure}[b]{0.33\textwidth}
        \centering
        \includegraphics[width=\textwidth]{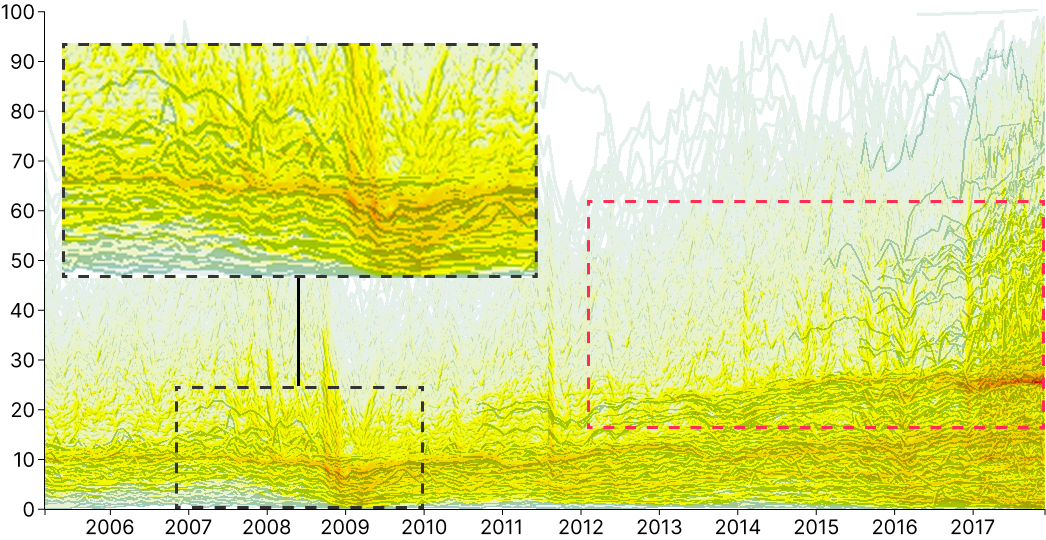}
        \caption{Our method when $\mu=0.25$, $\sigma=0.4$, $\eta=1.0$, $\phi=-25.0$}
        \label{fig:stock:b}
    \end{subfigure}
    \hfill
        \begin{subfigure}[b]{0.33\textwidth}
        \centering
        \includegraphics[width=\textwidth]{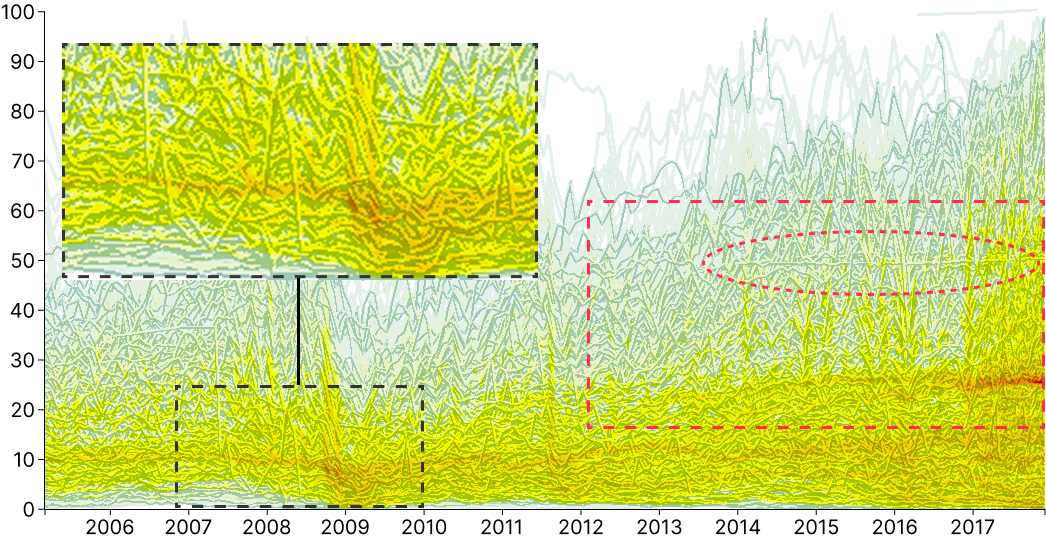}
        \caption{Our method when $\mu=0.6$, $\sigma=0.4$, $\eta=1.0$, $\phi=-25.0$}
        \label{fig:stock:c}
    \end{subfigure}\\
    \vspace{-3mm}
    \caption{Comparison of our method using a multi-hue colormap on the stock dataset~\cite{nyse}: (a) Direct application of the Lambertian model, the structure of the line field as well as outliers are not visible; (b) Our method with $\mu=0.25$, where main trends are emphasized;
    (c) Our method with $\mu=0.6$, which shows more outliers.
    A $3\times3$ kernel generates all of the results.
    }
    \label{fig:stock}
    \vspace{-1mm}
\end{figure*}

\begin{figure*}[tbp]
    \centering
    \hfill
       \begin{subfigure}[b]{0.31\textwidth}
        \centering
        \includegraphics[width=\textwidth]{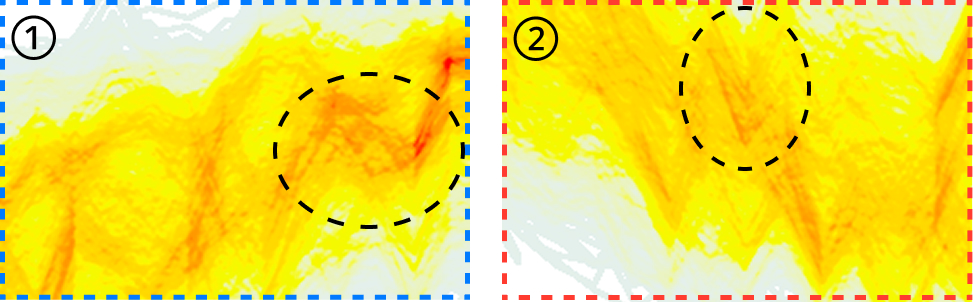}
        \label{fig:temperature:d}
    \end{subfigure}
    \hfill
    \begin{subfigure}[b]{0.31\textwidth}
        \centering
        \includegraphics[width=\textwidth]{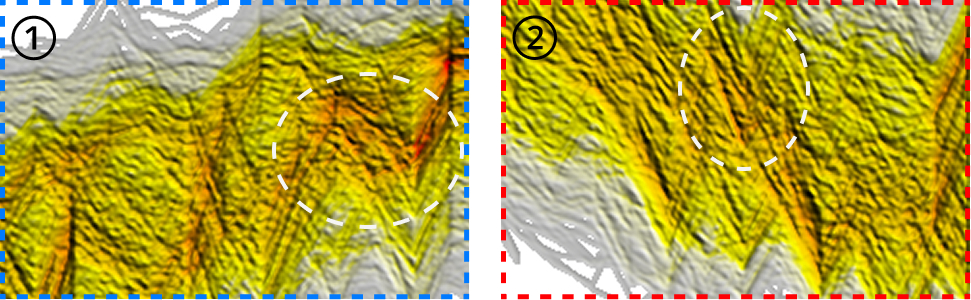}
        \label{fig:temperature:e}
    \end{subfigure}
    \hfill
    \begin{subfigure}[b]{0.31\textwidth}
        \centering
        \includegraphics[width=\textwidth]{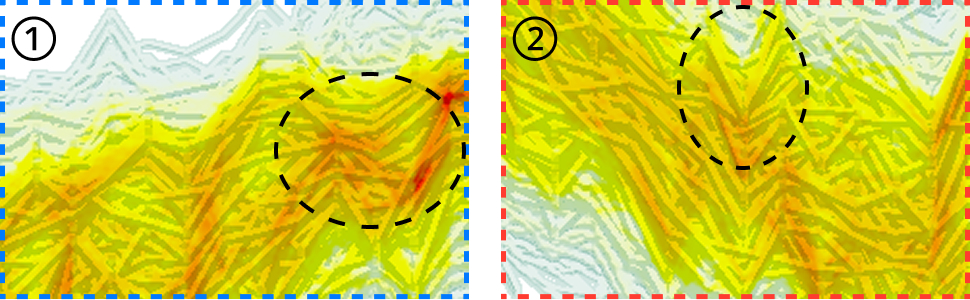}
        \label{fig:temperature:f}
    \end{subfigure}

    \vspace{-3mm}
    
    \begin{subfigure}[b]{0.33\textwidth}
        \centering
        \includegraphics[width=\textwidth]{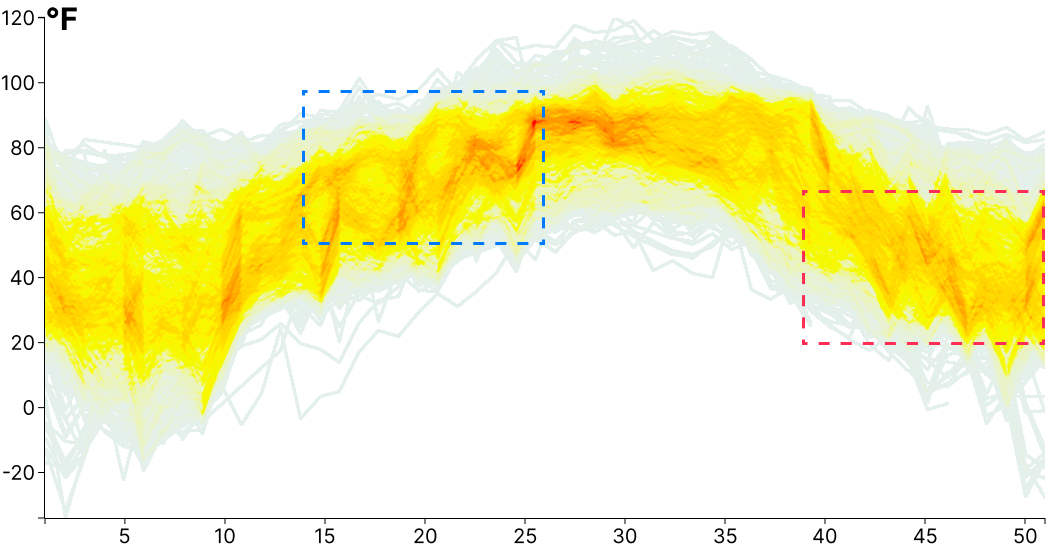}
        \caption{Plain density plot}
        \label{fig:temperature:a}
    \end{subfigure}
    \hfill
    \begin{subfigure}[b]{0.33\textwidth}
        \centering
        \includegraphics[width=\textwidth]{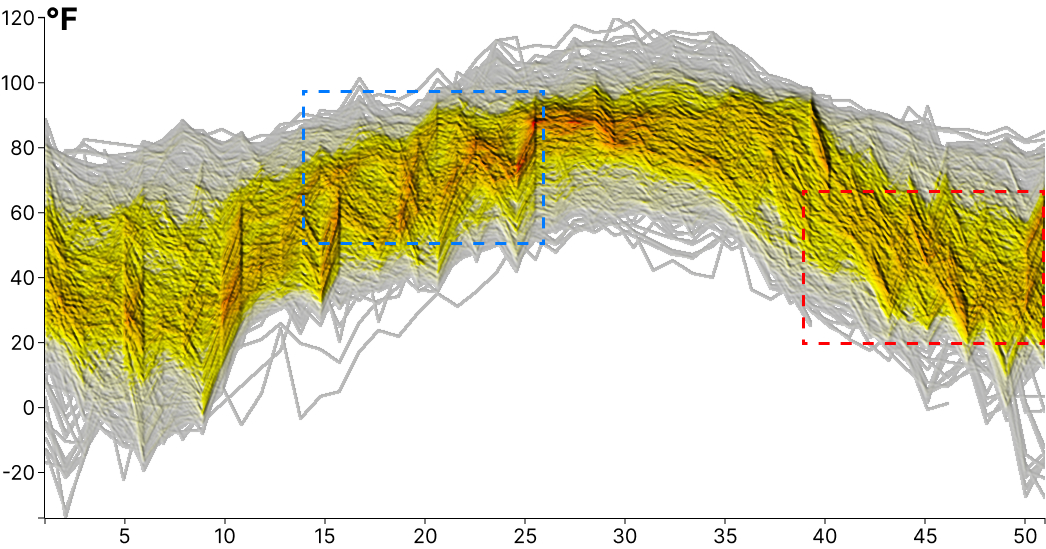}
        \caption{Direct Lambertian model}
        \label{fig:temperature:b}
    \end{subfigure}
    \hfill
    \begin{subfigure}[b]{0.33\textwidth}
        \centering
        \includegraphics[width=\textwidth]{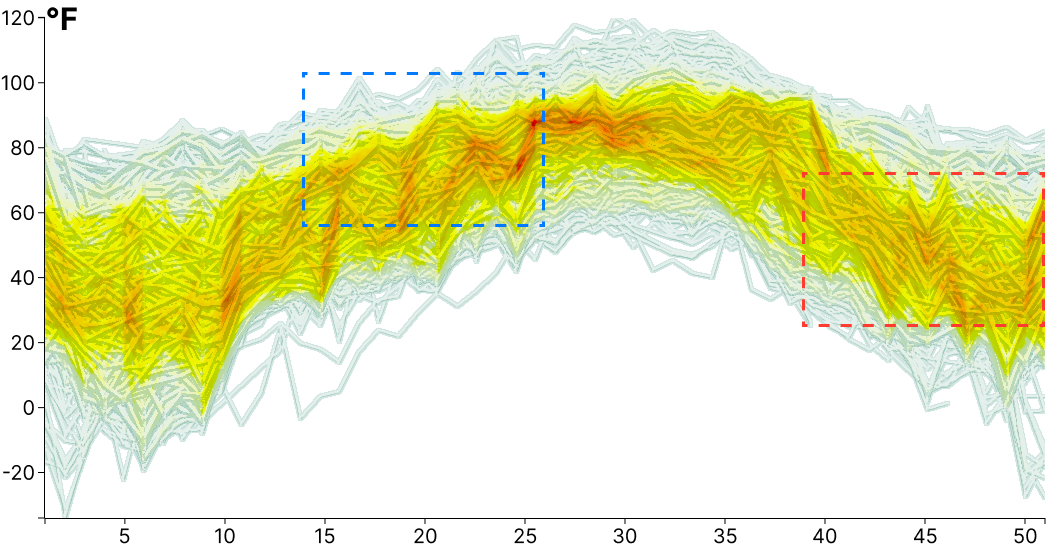}
        \caption{Our method when $\mu=0.1$, $\sigma=0.5$, $\eta=1.0$, $\phi=-20.0$}
        \label{fig:temperature:c}
    \end{subfigure}
    \hfill
    \vspace{-6mm}
    \caption{Application of our method on the temperature dataset~\cite{acis}. In the density plot of (a), the flow of the lines is not visible (see circled areas in cut-outs \ding{192} and \ding{193}), also not in the directly Lambertian shaded version of (b). Our method (c) allows the viewer to follow the ups and downs of the lines, which is especially visible in the encircled regions.
    A $5\times5$ kernel generates all of the results.}
    \label{fig:temperature}
    \vspace{-1mm}
\end{figure*}

\begin{figure*}[tbp]
    \centering
       \begin{subfigure}[b]{0.33\textwidth}
        \centering
        \includegraphics[width=\textwidth]{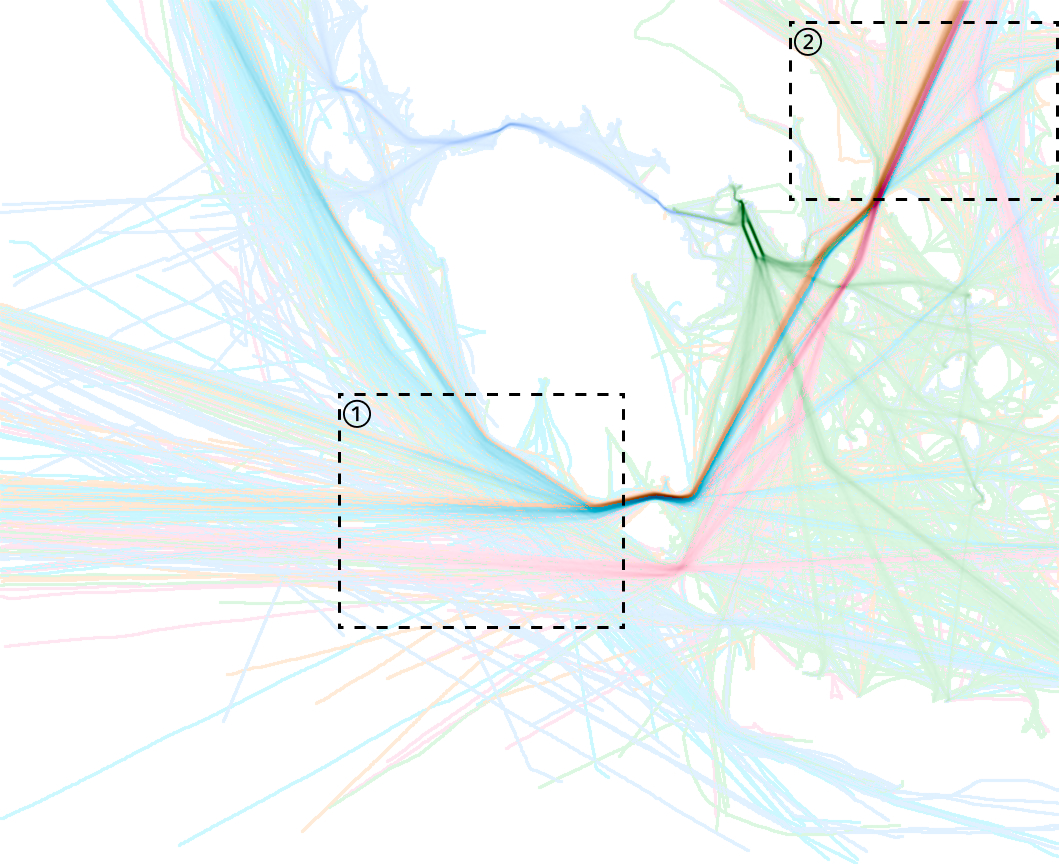}
        \caption{Plain density plot}
        \label{fig:ship_singlehue_raw}
    \end{subfigure}
    \hfill
    \begin{subfigure}[b]{0.33\textwidth}
        \centering
        \includegraphics[width=\textwidth]{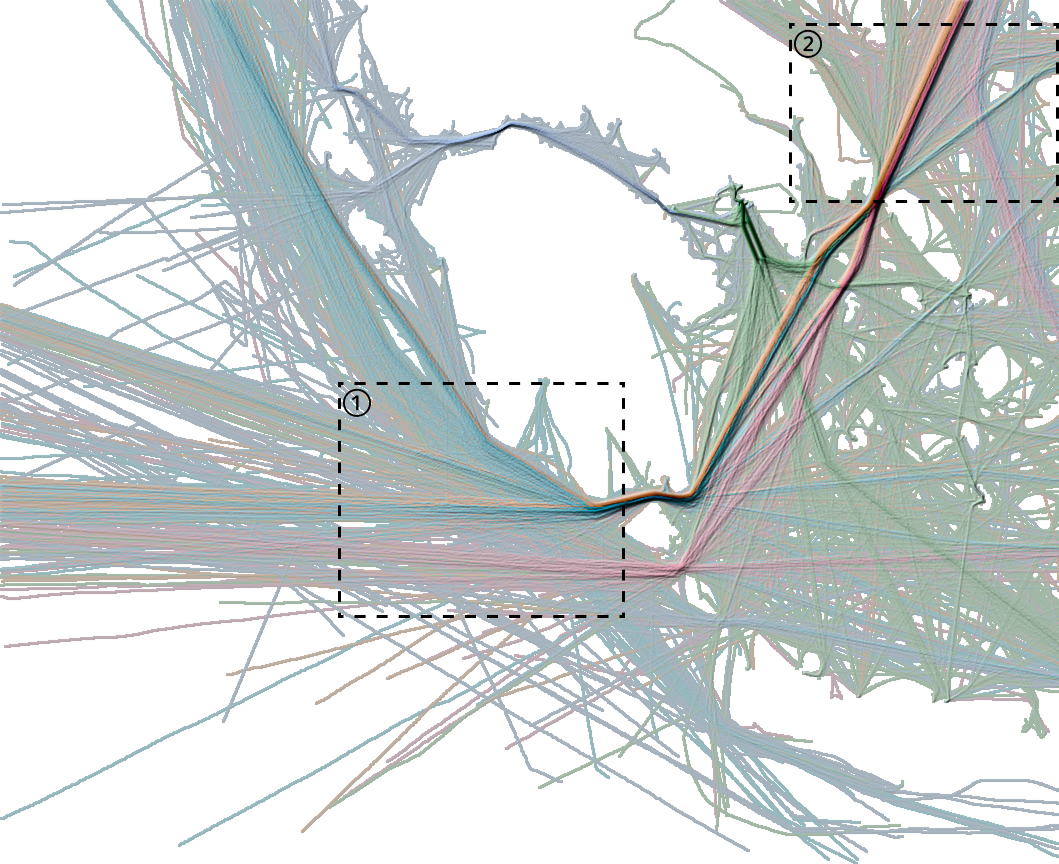}
        \caption{Direct Lambertian model}
        \label{fig:ship_singlehue_phong}
    \end{subfigure}
    \hfill
    \begin{subfigure}[b]{0.33\textwidth}
        \centering
        \includegraphics[width=\textwidth]{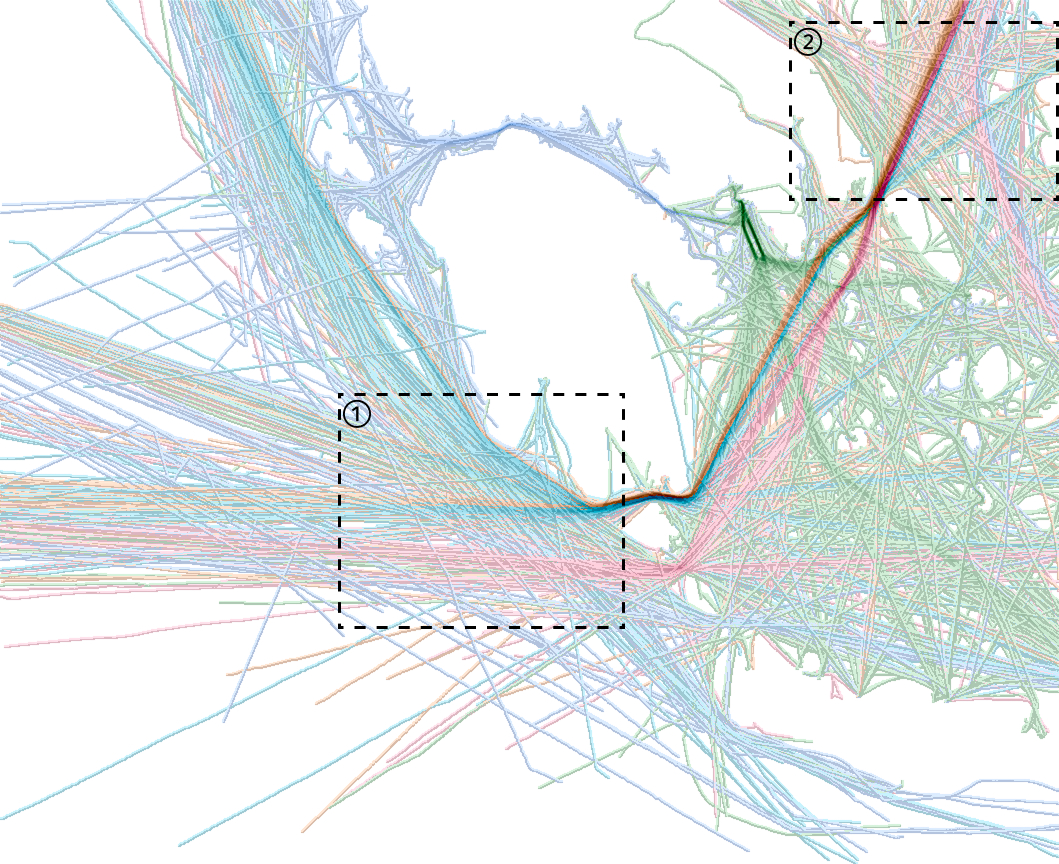}
        \caption{Our method when $\mu=1.0$, $\sigma=0.5$, $\eta=3.0$, $\phi=-20.0$}
        \label{fig:ship_singlehue_ours}
    \end{subfigure}
    \hfill
    \vspace{-6mm}
    \caption{Our method allows us to highlight differences between clusters. Here, we show the vessel trajectories with five clusters and hue-preserved shading. While direct Lambertian shading distorts the colors (b), our method combines good color fidelity with perceptual enhancement of lines crossing clusters in (c).
    A $3\times3$ kernel generates all of the results.}
    \label{fig:ship_singlehue}
\end{figure*}

\subsection{Case Study for Single Cluster}

\label{subsec:case_study}

So far, we have shown the result of our method using the vessel data set (Hellenic Trench AIS dataset) in \cref{fig:teaser}. Additionally, we utilize the NYSE stock trading dataset and the daily temperatures from the Applied Climate Information System (ACIS) Web Services to demonstrate our method. For all three datasets, we use a multi-hue color map to represent the density values. For reproducibility, the corresponding figure captions list all parameter settings (\(\mu, \sigma, \eta, \phi\)), and the discrete kernel size.

\noindent\textbf{Ship trajectories.}
The Hellenic Trench AIS dataset~\cite{frantzis2018hellenic} contains over 170{,}000 vessel trajectories collected to assess the impact of shipping routes on Mediterranean sperm whales~\cite{frantzis2019shipping}, from which 3{,}000 trajectories were sampled for visualization.
Direct diffuse (Lambertian) shading from density normals (see \cref{fig:teaser:b}) adds depth relative to the just density plot (\cref{fig:teaser:a}) but introduces chroma shifts and leaves many faint crossings unresolved.
All results in \cref{fig:teaser:c,fig:ship_multihue} use a \(3\times 3\) kernel with \(\eta=3.0\) and \(\phi=-20.0\).
In \cref{fig:teaser:c}, setting \(\mu=0.6\) and \(\sigma=0.5\) emphasizes trajectories of medium outlierness and injects a moderate amount of high-frequency detail, yielding coherent flow in high-density corridors (\textbf{D1}) while exposing fine structure in sparse regions (\textbf{D2}).
Varying \(\mu\) and \(\sigma\) provides user-steerable emphasis (\textbf{D3}). \cref{fig:ship_multihue} illustrates the influence of the parameter ranges on shading outcomes.
Focusing on inliers with \(\mu=0.0\) and \(\sigma=0.5\) (\cref{fig:ship_0.0_0.5_3.0_-20}) emphasizes high density patterns. In the boxed region~\ding{192}, the southwest--northeast corridor remains continuous despite intersecting northwest--southeast line trends, whereas \cref{fig:teaser:c} reverses the local emphasis between these crossings.
This capability is crucial as it allows analysts to interactively disambiguate dense intersections by selectively emphasizing the continuity of one trend over another, enabling task-driven visual exploration.
Conversely, emphasizing outliers with \(\mu=1.0\) and \(\sigma=0.5\) (\cref{fig:ship_1.0_0.5_3.0_-20}) highlights numerous deviant tracks that occlude the main trends in region~\ding{192}, making the underlying high-density routes difficult to recover from shading alone without the density colormap as a reference.
Under this setting, in region~\ding{193}, tracks crossing major shipping lanes are highlighted, while in region~\ding{194}, sparse connections involving ports on Crete (the southernmost endpoints in the region) are emphasized. This visually overshadows the much busier northwest--southeast routes within this region.
Reducing $\sigma$ to \(0.05\) with \(\mu=1.0\) (\cref{fig:ship_1.0_0.05_3.0_-20}) suppresses many outliers and re-illuminates the principal corridors across all marked regions; notably, in~\ding{194} a previously faint northwest--southeast pattern becomes discernible, albeit still partially masked by residual crossings.

\noindent\textbf{NYSE stock trading data.}
The NYSE dataset~\cite{nyse} contains 2{,}000 randomly sampled equities with closing prices in the 0--100 USD range from 2005--2017. \cref{fig:stock} compares a Lambertian diffuse baseline with two parameterizations of our method using a \(3\times 3\) kernel with \(\eta=1.0\) and \(\phi=-25.0\).
The density plot is noisy and contains many outliers, and exhibits two salient bands: a continuous low-value/low-volatility band near the bottom (black box) and a shorter band formed by post-2011 listings (at the bottom of the red box). Direct diffuse shading (\cref{fig:stock:a}) adds depth but introduces chroma shifts and rough textures in crowded regions, blurring dominant trends and masking sparse anomalies. With \(\mu=0.25\) and \(\sigma=0.4\) (\cref{fig:stock:b}), the bottom band becomes smoother and more contiguous in the black box, while in the red box, line details of the high-density pattern below become clear and short outlier traces appear above, both absent from the plain density and Lambertian renderings. Raising \(\mu\) to \(0.6\) but keeping \(\sigma\) fixed, (\cref{fig:stock:c}) exposes additional sparse structure and amplifies high-frequency detail because weak clustering and outlier-rich trajectories occupy more screen space. Our method also supports the inspection of individual trajectories and time series without losing context. E.g., in the black box, a timeseries with a sharp surge in 2008 and a drastic fall (American Lorain Corporation, ALN~\cite{aln}) can be seen.
In the red box, our method reveals a distinct, nearly flat trajectory (dashed red ellipse) that is lost in other views. This line is the preferred stock GLU-PRA~\cite{glua}, an atypical case whose price remains constant around \$50. Our visualization makes this unique, low-volatility pattern clearly visible for exploration.

\noindent\textbf{Temperature data.}
The ACIS dataset~\cite{acis} contains 293{,}175 weekly maximum temperature values from 6{,}187 U.S. time series; we visualize 2{,}000 randomly sampled ones. 
% Unlike the previous two examples where a \(3\times3\) kernel was used, 
Here, we adopt a \(5\times5\) discrete kernel to render slightly thicker lines. This reduces the number of simultaneously visible trajectories but improves the overall legibility of banded structures. In the plain density plot (\cref{fig:temperature:a}), heavy overlap limits trackable flow in the marked regions. Direct diffuse (Lambertian) shading (\cref{fig:temperature:b}) increases depth but perturbs the colormap and does not restore continuity. With our method configured at \(\mu=0.1\) and \(\sigma=0.5\) (\cref{fig:temperature:c}), the main seasonal trend is emphasized while a moderate amount of high-frequency detail is shown. In the encircled areas, most trajectories align with the dominant band yet reveal subtle local deviations that are not conveyed by the plain density or diffuse shading. Together, the larger kernel and the setting of \((\mu,\sigma)\) provide a clear, continuous visualization of the trend with selectively visible fine structure.

\subsection{Case Study with Multiple Clusters}
\label{sec:evaluation:single-hue}

To assess how our illumination model behaves under single-hue encodings, we revisit the Hellenic Trench AIS dataset and adopt the image-space colorization of Xue~\textit{et al.}~\cite{xue2024reducing} to assign distinct single-hue colormaps to five clusters. For a fair comparison to the multi-hue ship example in \cref{fig:teaser,fig:ship_multihue}, we reuse the same parameter setting with \cref{fig:ship_1.0_0.5_3.0_-20} that emphasized outliers there, i.e., we set a high outlier focus \(\mu=1.0\) and a half blend \(\sigma=0.5\), and keep \(\eta,\phi\) and the kernel size \(n\) identical (exact values are reported in the figure captions).

In the clustered density plot (\cref{fig:ship_singlehue_raw}), the dashed box highlights a region where multiple clusters interlace. Overlapping trajectories hinder the identification of local orientations and cross-cluster relationships. Applying direct diffuse (Lambertian) shading to the clustered density (\cref{fig:ship_singlehue_phong}) reveals some structure but introduces chroma shifts within single-hue palettes and suppresses fine inter-cluster crossings.

Our method (\cref{fig:ship_singlehue_ours}) highlights sparse, deviating trajectories that cross cluster boundaries, while ensuring the dominant trends within each cluster remain visually stable in their respective hues. Luminance-only composition preserves each cluster’s hue/chroma, preventing cross-hue contamination, and the structural normal map enhances transverse crossings and pronounced deviations with respect to major corridors, clarifying where flows traverse different colored clusters. This example intentionally emphasizes outlier revelation (\textbf{D2}) by using a higher $\mu$, and lowering \(\mu\) would instead favor within-cluster main-trend continuity (\textbf{D1}), illustrating user-steerable emphasis (\textbf{D3}) independent of the colormap strategy.

\section{Discussion and Conclusion} \label{sec:discussion}

Our method enhances line density plots by integrating a structural normal map derived from a trajectory outlierness measure with direction-aware illumination, thereby achieving continuity in high-density regions and revealing details in sparse regions. 

\noindent\textbf{Contributions and Broader Implications.} 
Our work makes two primary contributions: (1) a bin-based similarity metric to rank trajectories by shape congruence, and (2) a direction-aware, luminance-only illumination model that preserves colormap integrity. These components have broader applicability beyond our pipeline. The similarity metric itself (\cref{subsec:structure-enhancement}) can be used as a standalone analysis tool for data-driven focus+context or for querying trajectories based on shape (e.g., finding the ``most average'' or ``most anomalous'' path). The illumination model (\cref{subsec:illumination}) could be extended to other fields, such as scientific or medical visualization, where analysts must perceive complex line structures from large-scale data~\cite{eichelbaum13lineao}.

\noindent\textbf{Limitations and Future Directions.} 
Despite these contributions, several considerations remain.
Our proposed method requires four user parameters: $\mu$ and $\sigma$ for outlierness weighting, $\eta$ for normal-map scaling (\cref{eq:low_freq_normal}), and $\phi$ for L channel color composition. These enable flexible exploration but may overwhelm users unfamiliar with the data, since manual tuning is needed to balance trend visibility against outlier emphasis. For example, large $\mu$ values highlight outliers but risk overshadowing main trends in low-density regions, potentially leading to misinterpretation in static images (e.g., mistaking an outlier trajectory for a trend).
Similarly, normalizing outlierness scores to the $[0, 1]$ range for the $\mu$ slider makes the emphasis relative to the current dataset, which could visually exaggerate minor deviations in static views.
Interactive adjustment of $\mu$, as demonstrated in our video, alleviates this issue by enabling dynamic exploration. However, static visualizations could benefit from additional cues, such as distinct color encodings for trends versus outliers, to improve interpretability without user interaction. In addition, while working with our interface, we noticed that interactively moving light directions is a powerful mechanism to let selected clusters visually stand out. The moving light source creates a movement effect for the illuminated cluster, which visually separates it from all the other lines. We plan to explore this effect in future work.

A potential concern is that locally derived principal directions may produce salt-and-pepper noise if neighboring bins are assigned inconsistent orientations. In our datasets, such artifacts were not obvious because they contain coherent patterns, which is when density plots are most informative. However, in highly noisy data, adjacent bins could receive opposite light directions, leading to local visual noise. While we do not apply explicit smoothing, lightweight regularization strategies (e.g., spatial filtering of the direction field) could mitigate such artifacts in less structured datasets and remain a promising direction for future work.
% A potential concern is that locally derived principal directions may produce salt-and-pepper noise if neighboring bins are assigned inconsistent orientations. In our datasets, which contain coherent global or local patterns, such artifacts were not obvious. This is because density plots are most informative when trajectories exhibit meaningful structure; if data is dominated by noise, density visualization itself becomes less effective, and our method may also deteriorate. In such cases, adjacent bins could indeed receive nearly opposite light directions, leading to local visual noise. While our current implementation does not apply explicit smoothing, lightweight regularization strategies—such as spatial filtering of the direction field or consistency checks for neighboring orientations—could mitigate artifacts in less structured datasets. Exploring adaptive smoothing strategies tailored to local noise levels is a promising direction for future work.

As shown in \cref{subsec:performance} and \cref{fig:scalability}, scalability remains a key challenge with two distinct bottlenecks. The primary bottleneck is the outlierness computation (\cref{subsec:line-outlierness}), which is computationally intensive (e.g., 28.8 seconds for 10,000 lines on our test hardware). Although this is a one-time pre-processing step, its high cost is a significant limitation for rapid, iterative analysis of very large datasets.
A secondary bottleneck exists in the interactive parameter adjustment (e.g., $\mu$ and $\sigma$). In our current implementation, which is purely browser-based (front-end) and was tested on slightly older hardware (an Apple Silicon M1 laptop), we observe that adjusting parameters for datasets larger than 10,000 lines can cause delays exceeding 2 seconds, which hinders fluid exploration. As a bin-based method, the rendering performance also scales with the grid resolution of the density plot, creating a bottleneck for high-resolution displays.
Future work could address both limitations. The outlierness computation could be significantly accelerated using GPU-based parallel processing or approximate nearest-neighbor techniques. The interactive delay could be mitigated by (1) leveraging GPU acceleration (e.g., via WebGPU) for the front-end recalculation of normal maps and lighting, or (2) moving the intensive computations to a higher-performance backend server for rendering.

Finally, while our method enhances the visibility of outliers and structural continuity, the absence of formal user studies limits claims regarding perceptual effectiveness. Future work should include controlled evaluations to quantify how well trends and anomalies are distinguished, as well as the development of adaptive parameter-selection strategies to reduce user burden.

\noindent\textbf{Conclusion.} 
In summary, combining bin-based outlier control with direction-aware illumination constitutes a significant step forward in line density visualization, addressing long-standing challenges of continuity, detail enhancement, and color integrity. Despite the limitations discussed, future refinements---including direction-field smoothing, scalable computation, and perceptual validation---promise to establish this approach as a valuable tool for researchers and practitioners, enabling richer insights into complex line datasets.
%\input{sections/conclusion}

%% if specified like this the section will be ommitted in review mode
\acknowledgments{%
	This work was funded in part by Deutsche Forschungsgemeinschaft (DFG) Project 410883423 and Project 251654672 – TRR 161 ``Quantitative methods for visual computing.'' Yunhai Wang was supported by the grants of NSFC (No.62132017 and No.U2436209), the Shandong Provincial Natural Science Foundation (No.ZQ2022JQ32), the Fundamental Research Funds for the Central Universities, and the Research Funds of Renmin University of China.
}

\bibliographystyle{abbrv-doi-hyperref}

\bibliography{template}
\end{document}